\documentclass[lettersize,a4paper]{IEEEtran}
\usepackage[T1]{fontenc}
\usepackage{graphicx}

\usepackage{amsthm,amsmath,amssymb,bm}
\usepackage{mathrsfs}
\usepackage{amsfonts}
\newtheorem{theorem}{Theorem}

\newtheorem{lemma}{Lemma}

\newtheorem{remark}{Remark}
\newtheorem{assumption}{Assumption}

\usepackage{subfigure}
\usepackage{algorithm}
\usepackage{algpseudocode}
\usepackage{graphics}
\usepackage{epsfig}
\usepackage{cite}
\usepackage{color}
\usepackage{booktabs}
\usepackage{verbatim}

\begin{document}
\title{Joint Device Selection and Power Control for Wireless Federated Learning}
\author{Wei~Guo,~\IEEEmembership{Graduate Student Member,~IEEE}, Ran~Li,~\IEEEmembership{Graduate Student Member,~IEEE}, Chuan~Huang,~\IEEEmembership{Member,~IEEE}, Xiaoqi~Qin,~\IEEEmembership{Member,~IEEE}, Kaiming~Shen,~\IEEEmembership{Member,~IEEE}, and~Wei~Zhang,~\IEEEmembership{Fellow,~IEEE}
	\thanks{
		This work was supported in part the National Key R\&D Program of China with grant No. 2018YFB1800800, by Natural Science Foundation of China with grant No. 62022070, by the Guangdong Provincial Key Laboratory of Future Networks of Intelligence, by the Basic Research Project No. HZQB-KCZYZ-2021067 of Hetao Shenzhen-HK S\&T Cooperation Zone, and also by Shenzhen Science \& Innovation Fund under Grant JCYJ20180507182451820, and the Australian Research Council’s Project funding scheme under LP160101244. This work was submitted in part to 2022 IEEE Global Communications Conference. ({\it Corresponding author: Chuan Huang.})
		
		W. Guo and R. Li are with the School of Science and Engineering (SSE) and Future Network of Intelligence Institute (FNii), The Chinese University of Hong Kong, Shenzhen, China, 518172. Emails: weiguo1@link.cuhk.edu.cn and ranli2@link.cuhk.edu.cn. 
		
		C. Huang and K. Shen are currently with the School of Science and Engineering (SSE) and Future Network of Intelligence Institute (FNii), The Chinese University of Hong Kong, Shenzhen, China, 518172, and Peng Cheng Laboratory, Shenzhen, China, 518066. Emails: huangchuan@cuhk.edu.cn and shenkaiming@cuhk.edu.cn.
		
		X. Qin is with the State Key Laboratory of Networking and Switching Technology, Beijing University of Posts and Telecommunications, Beijing, China, 100876. Emial: xiaoqiqin@bupt.edu.cn.
		
		W. Zhang is with the School of Electrical Engineering and Telecommunications, University of New South Wales, Sydney, Australia, NSW 2052. Email: w.zhang@unsw.edu.au.}}

\maketitle

\begin{abstract}
	This paper studies the joint device selection and power control scheme for wireless federated learning (FL), considering both the downlink and uplink communications between the parameter server (PS) and the terminal devices. In each round of model training, the PS first broadcasts the global model to the terminal devices in an analog fashion, and then the terminal devices perform local training and upload the updated model parameters to the PS via over-the-air computation (AirComp). First, we propose an AirComp-based adaptive reweighing scheme for the aggregation of local updated models, where the model aggregation weights are directly determined by the uplink transmit power values of the selected devices and which enables the joint learning and communication optimization simply by the device selection and power control. Furthermore, we provide a convergence analysis for the proposed wireless FL algorithm and the upper bound on the expected optimality gap between the expected and optimal global loss values is derived. With instantaneous channel state information (CSI), we formulate the optimality gap minimization problems under both the individual and sum uplink transmit power constraints, respectively, which are shown to be solved by the semidefinite programming (SDR) technique. Numerical results reveal that our proposed wireless FL algorithm achieves close to the best performance by using the ideal {\it FedAvg} scheme with error-free model exchange and full device participation.
\end{abstract}


\IEEEpeerreviewmaketitle

\section{Introduction}\label{sec:intro}
The future 6th generation (6G) wireless communication systems are envisioned to support various intelligent applications and services, which are empowered by the significant increase of wireless edge devices (e.g., mobile phones and sensors) with growing computation and communication capabilities \cite{KBLetaief}. A vast amount of data generated by the edge devices can be utilized to train machine learning (ML) models to further enhance the intelligence of those applications and services \cite{MChiang}. Conventional machine learning methods, particularly those based on deep neural networks (DNN), are centralized and require to collect all the raw data to the central server or cloud in order to train the artificial intelligence (AI) model \cite{DL_book}. However, certain privacy and security issues arise during the data migrations, and limited wireless resource also poses some new technical challenges for the design of wireless AI systems \cite{CXJiang,CZhang}.

To address the issues of the centralized ML, federated learning (FL), a new distributed ML paradigm, was proposed in \cite{HBMcMahan}, which achieves great success and becomes increasingly popular in both the academia and industry. In FL, the distributed terminal devices, orchestrated by a single parameter server (PS), collaboratively train an AI model in an iterative fashion. During the whole  iterative process, all participated terminal devices only exchange the model parameters with the PS and keep the raw data locally to protect the privacy and security. Nonetheless, FL faces several new challenges when being implemented in the wireless scenarios: $1)$ data heterogeneity: different from the centralized ML, FL highly suffers from the data heterogeneity \cite{JKone,WYBLim}, which arises when the distributions of the generated data vary from device to device; and $2)$ communication complexity: the total training process for FL consumes a large amount of communication resources for serving large bunches of terminal devices, since the desired ML model is in general of high dimension and the number of updating rounds in the training process is thus considerably large \cite{YZhao,TLi}.

Aimed at enhancing the communication efficiency of the FL, comprehensive studies have been done from a communication perspective \cite{MChen,JXu,WShi, MMAmiri_2,YWang, SWang, KYang, YSun, NZhang,GZhu,GZhu_1,XCao,XFan}. The authors in \cite{MChen} proposed a joint communication and learning framework and formulated an optimization problem considering the user selection and wireless resource allocation to minimize the FL training loss. In \cite{JXu}, the authors identified the temporal dependency and varying significance of the training rounds, and proposed a joint device selection and bandwidth allocation scheme to maximize the weighted sum of selected clients in the long-term view. In \cite{MMAmiri_2}, the authors proposed an update-aware device scheduling and resource allocation policy and analyzed the convergence of the FL algorithm with device scheduling. In \cite{WShi}, under the latency constraint, the authors proposed a joint device scheduling and resource allocation scheme to minimize the global loss of wireless FL, which achieves a desirable trade-off between the number of required training rounds and the latency per round. In \cite{YWang}, the authors proposed a joint wireless resource and quantization bits allocation policy to minimize the quantization error while guaranteeing the transmission outage probability of the uplink transmissions for the wireless FL. In \cite{SWang}, the authors proposed a control algorithm to determine the learning parameter to minimize the optimality gap between the expected and optimal loss function values. However, all the aforementioned works are conducted under the digital communications, and consequently the communication overhead and latency still increase with the number of participated edge devices \cite{GZhu}. Recently, by investigating the waveform superposition nature of the wireless multiple access (MAC) channels \cite{BNazer}, over-the-air computation (AirComp) enabled analog uplink transmission for model aggregation methods have been proposed for the FL \cite{KYang,YSun, NZhang,GZhu,GZhu_1,XCao,XFan}. Specifically, in \cite{KYang}, the authors proposed a joint device scheduling and beamforming design to minimize the mean square error (MSE) of the aggregated signals to accelerate the convergence of the AirComp FL. In \cite{YSun}, the authors proposed an energy-aware device scheduling algorithm under the energy constraint to optimize the AirComp FL performance. In \cite{GZhu}, the authors proposed a broadband AirComp FL and investigated the corresponding power control and device scheduling problem, and the authors in \cite{GZhu_1} further adopts a one-bit quantization scheme to modify the policy adopted in \cite{GZhu}. The authors in \cite{NZhang} proposed a gradient aware power control scheme to enhance the performance of the AirComp FL. In \cite{XCao}, the authors proposed a power control to minimize the optimality gap between the expected and optimal global loss values of the AirComp FL, and parallelly in \cite{XFan}, the authors proposed a joint device selection and power control for the AirComp FL with uniform power scaling and equal weights model aggregation.

All the above researches about FL over wireless channels only considered the uplink transmissions for the model parameters from terminal devices to the PS, assuming the availability of accurate global model at the devices through perfect downlink transmissions. However, the downlink transmissions usually suffer from quantization error, limited transmit power, limited bandwidth, and additive noise, which degrades the performance of the considered FL systems \cite{SCaldas, XWei, MMAmiri,MMAmiri_1,JHAhn}. Therefore, broadcasting of inaccurate global model is also an important issue for the implementation of the FL algorithms. Without considering the wireless channels, the authors in \cite{SCaldas} proposed a linear projection method to broadcast a compressed global model to the  devices, and the authors in \cite{XWei} derived the sufficient conditions for controlling the signal-to-noise ratios (SNRs) of both the downlink and uplink transmission to maintain the linear convergence rate of the FL algorithm \cite{XWei}. Different from \cite{SCaldas,XWei} and under wireless implementations of FL, the authors in \cite{MMAmiri,MMAmiri_1} provided the convergence analysis of both the digital and analog downlink transmissions and showed the advantages of the analog downlink transmission, and the same result was presented in \cite{JHAhn} numerically.  To the best of our knowledge, there are few existing works addressing how to efficiently mitigate the impact of both the downlink and uplink wireless communications on the FL algorithm.

This paper provides a comprehensive analysis on the wireless FL algorithm, considering the analog (uncoded) downlink transmissions for global model broadcasting and AirComp enabled analog uplink transmissions for model aggregation, due to the superiority of the analog downlink and uplink transmissions for wireless FL as mentioned above. Compared to the previous works that only account for the uplink communications of FL \cite{KYang,YSun,NZhang,GZhu,GZhu_1,XCao,XFan} and only focus on the convergence analysis for FL \cite{XWei, MMAmiri,MMAmiri_1}, the goals of this paper are not only to investigate the impact of both the downlink and uplink wireless communications on the convergence behavior of the considered FL algorithm, but also to mitigate this impact to enhance the FL performance. The main contribution of this paper is summarized as follows:
\begin{itemize}
	\item We propose an AirComp-based adaptive reweighing scheme for model aggregation, where the adaptive weights are directly determined by the uplink transmit power values of the participated devices in each communication round. With respect to our proposed policy, we analyze the convergence behavior of the wireless FL algorithm and derive the upper bound on the expected optimality gap between the expected and optimal global loss values, which determines how bias the FL converges and theoretically quantifies the impact of both the downlink and uplink wireless communications on the convergence of the wireless FL algorithm, and thus needs to be minimized to enhance the FL performance.
	\item  With only instantaneous channel state information (CSI) known per round of the learning process, we formulate the optimality gap minimization problem as optimizing over the device selection and power control round by round, under both the individual and sum uplink transmit power constraints, respectively. Though the optimality gap minimization problem is a mixed integer programming (MIP) problem, we transform it into a continuous quadratically constrained ratio of two quadratic functions (QCRQ) minimization problem, which is efficiently solved by the semidefinite relaxation (SDR) technique. Furthermore, based on the optimal SDR solution, we derive the optimal device selection and power control to minimize the optimality gap for both the individual and sum uplink power constraint cases.
\end{itemize}

The reminder of this article is organized as follows. Section \ref{sec:system} presents the system model. In Section \ref{sec:convergence}, we provide the convergence analysis results of the wireless FL system in terms of the optimality gap. In Section \ref{sec:gap_min}, we formulate the optimality gap minimization problem and find the optimal device selection and power control to minimize the optimality gap. Numerical results are presented in Section \ref{sec:numerical}. Finally, Section \ref{sec:conclusion} concludes this paper. 

{\it Notation:} The bold lower-case letter denotes a vector; the bold upper-case letter denotes a matrix; the calligraphic upper-case letter denotes a set. $\mathbb{E}(\cdot)$ denotes the expectation operator. $\nabla$ denotes the gradient operator. For a vector $\mathbf{x}$, $\|\mathbf{x}\|_2$ denotes the Euclidean norm of $\mathbf{x}$; $\mathbf{x}^T$ denotes the transpose of $\mathbf{x}$; $\mathbf{x}^H$ denotes the Hermitian transpose of a complex vector $\mathbf{x}$. $\langle\mathbf{x},\mathbf{y}\rangle$ denotes the inner product between vectors $\mathbf{x}$ and $\mathbf{y}$. $\mathbf{0}$ and $\mathbf{1}$ denote the null vector and all-one vector, respectively. For a matrix $\mathbf{M}$, $\mathbf{M}^T$, $\mathtt{tr}(\mathbf{M})$, and $\mathtt{rank}(\mathbf{M})$ denote the transpose, trace, and rank, of $\mathbf{M}$, respectively. $[\mathbf{M}]_{i,j}$ denotes the $(i,j)$-th element of $\mathbf{M}$. $\mathtt{diag}(x_1,\cdots,x_n)$ denotes a diagonal matrix with $x_1,\cdots,x_n$ being its diagonal elements. $\mathbf{M}\succeq 0$ means that matrix $\mathbf{M}$ is positive semidefinite. $\mathbf{I}$ denotes the identity matrix. $\mathbb{R}$ and $\mathbb{C}$ denote the sets of real numbers and complex numbers, respectively. We denote a circularly symmetric complex Gaussian (CSCG) distribution with the real and imaginary components with variance $\sigma^2/2$ by $\mathcal{CN}(0,\sigma^2)$. For a set $\mathcal{A}$, $|\mathcal{A}|$ denotes its cardinality.

\section{System Model}\label{sec:system}
\subsection{Preliminaries}\label{subsec:preliminaries}
\begin{figure}[H] 
	\centering 
	\includegraphics[width=0.45\textwidth]{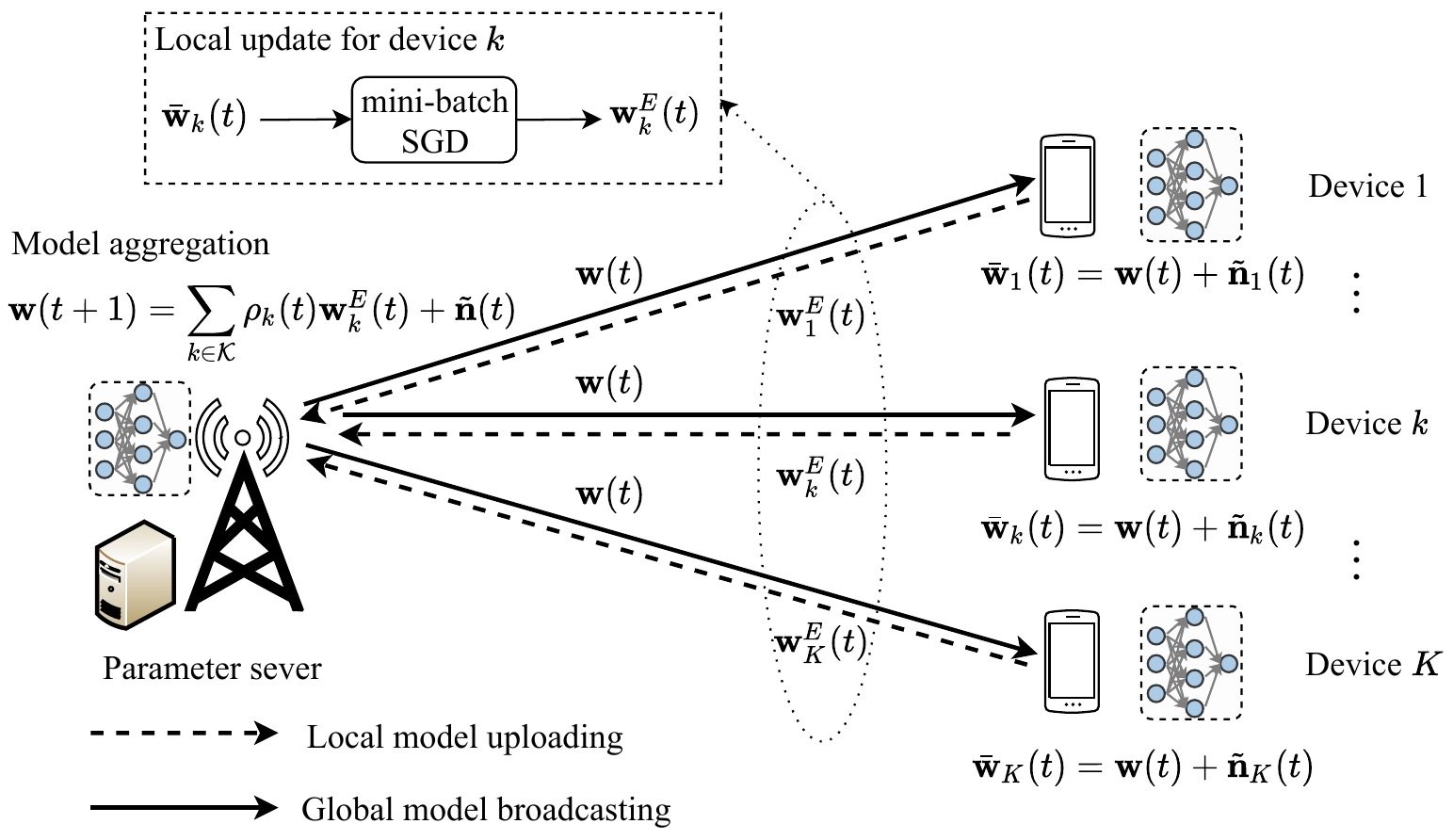} 
	\caption{Wireless FL system with one PS and $K$ terminal devices.}
	\label{fig:system}
\end{figure}
We consider a wireless FL system, as shown in Fig. \ref{fig:system}, where one PS coordinates a set  of $K$ terminal devices $\mathcal{K}=\{1,\cdots,K\}$ through wireless channels to cooperatively train a shared ML model (e.g., DNN), denoted by $\mathbf{w}\in\mathbb{R}^d$  of dimension $d$. Each terminal device $k\in\mathcal{K}$ collects its own labeled training data, and constitutes a local data set $\mathcal{D}_k=\{(\mathbf{x}_{k,i},y_{k,i})\}_{i=1}^{D_k}$ with $D_k=|\mathcal{D}_k|$ data samples, where $\mathbf{x}_{k,i}\in\mathbb{R}^n$ is the $i$-th input data with $n$ dimensions and $y_{k,i}\in\mathbb{R}$ is the labeled output corresponding to $\mathbf{x}_{k,i}$. Then, the total training data set is given as $\mathcal{D}=\bigcup_{k\in\mathcal{K}}\mathcal{D}_k$ of size $D=|\mathcal{D}|=\sum_{k=1}^KD_k$.

Though the PS has no access to the data samples distributed at the terminal devices due to the privacy concern, the goal of training a global model $\mathbf{w}$ can be achieved by solving the following distributed learning problem
\begin{align}\label{FL_problem}
	\min_{\mathbf{w}}\ F(\mathbf{w})=\sum_{k=1}^Kq_kF_k(\mathbf{w}),
\end{align}
where $q_k=D_k/D$ is the fraction of data samples for device $k$ and $F_k(\mathbf{w})$ is the local loss function defined as 
\begin{equation}\label{eq:local_loss}
	F_k(\mathbf{w})=\frac{1}{D_k}\sum_{i=1}^{D_k}\mathcal{L}(\mathbf{w};(\mathbf{x}_{k,i},y_{k,i})),
\end{equation}
with $\mathcal{L}(\mathbf{w};(\mathbf{x}_{k,i},y_{k,i}))$ being an empirical sample-wise loss function defined by the learning task that quantifies the loss of the model $\mathbf{w}$ for sample $(\mathbf{x}_{k,i},y_{k,i})$.

For the implementation of FL algorithms, the system solves the distributed learning problem \eqref{FL_problem} in an iterative fashion following the widely used {\it broadcasting-computation-aggregation} (BCA) principle, which involves the following three steps in each iteration: $1)$ the PS broadcasts a global model $\mathbf{w}$ to the terminal devices; $2$) terminal device $k$ updates the global model $\mathbf{w}$ to a local model $\mathbf{w}_k\in\mathbb{R}^d$ after several local learning iterations based on its local data samples; $3)$ the PS aggregates $\mathbf{w}_k$'s to generate a new global model $\mathbf{w}$. In the following, we discuss the considered wireless FL algorithm.

\subsection{Wireless FL Algorithm}\label{subsec:wireless_FL}
This paper considers the scenario where both the downlink and uplink communications are performed in the analog manner and the quasi-static fading channel model is adopted, i.e., both the downlink and uplink channels remain unchanged during each communication round (corresponding to one FL iteration containing all the three BCA steps) and are independent across different communication rounds following the Rayleigh distribution. Specifically, at the $t$-th communication round, the three steps of the wireless FL algorithm are described as follows:
\begin{itemize}
	\item[(1)]{\textbf{Broadcasting:}} In this step, the PS broadcasts the vector $\mathbf{u}_{\rm dl}(t)=p_s(t)\mathbf{w}(t)$ to terminal devices, containing the information of the global model $\mathbf{w}(t)$ and with the downlink transmit power value $p_s(t)$ satisfying
	\begin{equation}\label{eq:dl_power_budget}
		\|\mathbf{u}_{\rm dl}(t)\|_2^2=\|p_s(t)\mathbf{w}(t)\|_2^2\leq P_{\rm max}^{\rm dl}(t),
	\end{equation}
	where $P_{\rm max}^{\rm dl}(t)$ is the downlink transmit power budget at the $t$-the round. Hence, the received vector at the $k$-th terminal device is given as
	\begin{equation}\label{eq:rx_model}
		\begin{aligned}
			\mathbf{v}_k(t)&=h_k^{\rm dl}(t)\mathbf{u}_{\rm dl}(t)+\mathbf{n}_k(t)\\
			&=h_k^{\rm dl}(t)p_s(t)\mathbf{w}(t)+\mathbf{n}_k(t),
		\end{aligned}
	\end{equation}
	where $h_k^{\rm dl}\in\mathbb{C}$ is the complex-valued downlink channel coefficient between the PS and terminal device $k$, and $\mathbf{n}_k(t)\in\mathbb{C}^d$ is the independent identically distributed (i.i.d.) CSCG noise vector following the distribution $\mathcal{CN}(0,\sigma_d^2\mathbf{I})$ with $\sigma_d^2$ being the downlink noise power. With perfect knowledge of $p_s(t)$ and the CSI of the downlink channels, terminal device $k$ estimates the global model by descaling the received signal $\mathbf{v}_k$, i.e.,
	\begin{equation}\label{eq:estimate_model}
		\bar{\mathbf{w}}_k(t)=\frac{a_k(t)}{p_s(t)h_k^{\rm dl}(t)}\mathbf{v}_k(t)=\mathbf{w}(t)+\tilde{\mathbf{n}}_k(t),
	\end{equation}
	where $\bar{\mathbf{w}}_k(t)$ is the estimated global model for terminal device $k$ at the $t$-th round, $a_k(t)$ is the selection indicator for the device $k$ at the $t$-th round ($a_k(t)=1$ implies that the device $k$ is selected to participate in the training at the $t$-th round, and otherwise, $a_k(t)=0$), and $\tilde{\mathbf{n}}_k(t)=\frac{a_k(t)}{p_s(t)h_k^{\rm dl}(t)}\mathbf{n}_k(t)$ is the equivalent noise at the device $k$. Then, terminal device $k$ sets its current local model to $\mathbf{w}_k(t)=\bar{\mathbf{w}}(t)$.
	\item[(2)]{\textbf{Local Model Update:}} In this step, terminal device $k$ updates its local model $\mathbf{w}_k(t)$ by minimizing its local objective \eqref{eq:local_loss} via a local optimization algorithm, e.g., the mini-batch stochastic gradient descent (SGD) method, based on its collected data set $\mathcal{D}_k$. When implementing the mini-batch SGD algorithm, the local dataset $\mathcal{D}_k$ at the device $k$ is divided into several mini-batch with size $B$, and the device runs $E$ SGD iterations with each SGD iteration corresponding to one mini-batch. At the $\tau$-th SGD iteration, $\tau\in\{1,\cdots,E\}$, the local model is updated as
	\begin{equation}\label{eq:local_sgd}
		\mathbf{w}_k^{\tau}(t)=\mathbf{w}_k^{\tau-1}(t)-\eta_t\nabla F_k(\mathbf{w}_k^{\tau-1}(t);\mathcal{B}_k^{\tau}(t)),
	\end{equation}
	where $\eta_t$ is the learning rate at the $t$-th round, $\mathcal{B}_k^{\tau}(t)$ is a mini-batch with size $|\mathcal{B}_k^{\tau}(t)|=B$ and its data points being independently and uniformly chosen from $\mathcal{D}_k$, and $\nabla F_k(\mathbf{w}_k^{\tau-1}(t);\mathcal{B}_k^{\tau}(t))\in\mathbb{R}^{d}$ is the stochastic gradient of the local loss function with respect to (w.r.t.) the $(\tau-1)$-th model parameter $\mathbf{w}_k^{\tau-1}(t)$ and randomly sampled mini-batch $\mathcal{B}_k^{\tau}(t)$, i.e.,
	\begin{align}\label{eq:local_sgd_gradient}
		\nabla F_k&(\mathbf{w}_k^{\tau-1}(t);\mathcal{B}_k^{\tau}(t))\\\nonumber
		&=\frac{1}{B}\sum_{\xi_{k,i}^{\tau}(t)\in\mathcal{B}_k^{\tau}(t)}\nabla\mathcal{L}(\mathbf{w}_k^{\tau-1}(t);\xi_{k,i}^{\tau}(t)),
	\end{align}
	with $\xi_{k,i}^{\tau}(t)$ being the $i$-th training data in mini-batch $\mathcal{B}_k^{\tau}(t)$. The initial local model parameter at device $k$ before training is set as $\bar{\mathbf{w}}(t)$ in \eqref{eq:estimate_model}, i.e., $\mathbf{w}_k^{0}(t)=\bar{\mathbf{w}}_k(t)$. After $E$ SGD iterations, the local model parameter at device $k$ is updated as $\mathbf{w}_k^E(t)$.
	\item[(3)]{\textbf{Model Aggregation:}} In this step, the terminal devices upload their updated local model parameters $\mathbf{w}_k^E(t)$'s to the PS, and the PS aggregates the received local models to generate a new global model. For the uplink transmissions of $\mathbf{w}_k^E(t)$'s, the terminal devices transmit their local model parameters concurrently through AirComp by exploiting the waveform superposition nature of the wireless MAC channel, since the information of interest at the PS for aggregation is just the weighted summation of the local model parameters. Specifically, the updated model parameter $\mathbf{w}_k^E(t)$ at the device $k$ is multiplied with a pre-processing factor $\beta_k(t)$, which is given as
	\begin{equation}\label{eq:processing_factor}
		\beta_k(t)=\frac{a_k(t)p_k(t)(h_k^{\rm up}(t))^H}{|h_k^{\rm up}(t)|^2},
	\end{equation}
	where $p_k(t)$ is the uplink transmit power value at device $k$, $h_k^{\rm up}(t)$ is the complex uplink channel coefficient from device $k$ to the PS. With perfect CSI of the uplink channels, the received vector $\mathbf{v}(t)$ at the PS is given as
	\begin{equation}\label{eq:server_rx_model}
		\begin{aligned}
			\mathbf{v}(t)&=\sum_{k\in\mathcal{K}}h_k^{\rm up}(t)\beta_k(t)\mathbf{w}_k^{E}(t)+\mathbf{n}(t)\\
			&=\sum_{k\in\mathcal{K}}a_k(t)p_k(t)\mathbf{w}_k^{E}(t)+\mathbf{n}(t),
		\end{aligned}		
	\end{equation} 
	where $\mathbf{n}(t)\in\mathbb{C}^d$ is the i.i.d. CSCG noise vector following the distribution $\mathcal{CN}(0,\sigma_u^2\mathbf{I})$. In this paper, the uplink transmit power values at the devices satisfy the following two types of constraints: For the individual uplink transmit power constraint at each device, the transmit power value of device $k$ is supposed to satisfy
	\begin{equation}\label{eq:individual_power_constraint}
		\|\beta_k(t)\mathbf{w}_k^{E}(t)\|_2^2\leq P_{\rm max}^k(t),
	\end{equation}
	where $P_{\rm max}^k(t)$ is the individual power budget at the $t$-th round of device $k$; for the sum uplink transmit power constraint over all devices, the transmit power values of the devices are supposed to satisfy
	\begin{equation}\label{eq:sum_power_constraint}
		\sum_{k=1}^K\|\beta_k(t)\mathbf{w}_k^{E}(t)\|_2^2\leq P_{\rm tot}(t),
	\end{equation}
	where $P_{\rm tot}(t)$ is the sum power budget at the $t$-th round of all the participated devices.
	
	The PS scales the received signal $\mathbf{v}(t)$ with a factor $1/\zeta(t)$ to aggregate and update the global model parameter as
	\begin{equation}\label{eq:model_aggregation}
		\begin{aligned}
			&\mathbf{w}(t+1)\\
			=&\frac{\mathbf{v}(t)}{\zeta(t)}=\frac{1}{\zeta(t)}\sum_{k\in\mathcal{K}}a_k(t)p_k(t)\mathbf{w}_k^{E}(t)+\frac{1}{\zeta(t)}\mathbf{n}(t),
		\end{aligned}
	\end{equation}
	where $\zeta(t)$ is set as  the summation of all the products $a_k(t)p_k(t)$, i.e., $\zeta(t)=\sum_{k=1}^Ka_k(t)p_k(t)$, with perfect knowledge on all $a_k(t)$ and $p_k(t)$ at the PS. Hence, based on \eqref{eq:server_rx_model} and \eqref{eq:model_aggregation}, the model aggregation is given as\footnote{The proposed model aggregation scheme requires perfect CSI about both the downlink and uplink channels. Nevertheless, our design and analysis can be extended to the case with imperfect CSI, and the impact of imperfect CSI will be studied for future work.}
	\begin{equation}\label{eq:propose_aggregation}
		\mathbf{w}(t+1)=\sum_{k\in\mathcal{K}}\rho_k(t)\mathbf{w}_k^{E}(t)+\tilde{\mathbf{n}}(t),
	\end{equation}
	where $\rho_k(t)=\frac{a_k(t)p_k(t)}{\sum_{j\in\mathcal{K}}a_j(t)p_j(t)}$ is the weight of $\mathbf{w}_k^{E}(t)$ for aggregation satisfying $\sum_{k=1}^K\rho_k(t)=1$ and $\tilde{\mathbf{n}}(t)=\frac{\mathbf{n}(t)}{\sum_{j\in\mathcal{K}}a_j(t)p_j(t)}$ is the equivalent noise.
\end{itemize}

\begin{remark}\label{rem:sel_motivation}
	Due to the dynamic nature of the wireless channels, the terminal devices may encounter a relatively weak downlink channel, i.e., $|h_k^{\rm dl}(t)|\approx 0$, when downloading the global model from the PS. Then, the received model $\bar{\mathbf{w}}_k(t)$ at device $k$ will be severely damaged by the equivalent noise $\tilde{\mathbf{n}}_k(t)$, as shown in \eqref{eq:estimate_model}. Similarly, the devices may also encounter a relatively weak uplink channel, i.e., $|h_k^{\rm up}(t)|\approx 0$, when uploading the updated model $\mathbf{w}_k^E(t)$ to the PS. Then, the device $k$ may not be capable of transmitting $\mathbf{w}_k^E(t)$ due to the limited transmit power budget according to \eqref{eq:processing_factor}, \eqref{eq:individual_power_constraint}, and \eqref{eq:sum_power_constraint}. Thus, device selection according to the dynamic uplink and downlink channel conditions under the limited transmit power budget is necessary for the wireless FL systems.
\end{remark}

As mentioned in Remark \ref{rem:sel_motivation}, when practically implementing the wireless FL algorithm, not all devices are able to participate the training over the whole time. Hence, when considering both the downlink and uplink communications between the PS and the terminal devices, we need to develop an efficient device selection scheme while guaranteeing the convergence of the considered algorithm. 

\begin{remark}\label{rem:joint_sel_power}
	In our proposed weighted model aggregation scheme \eqref{eq:propose_aggregation}, under the wireless environment, the adaptive weights in each round are determined directly by the uplink transmit power values of the participated terminal devices, which are controllable and can be optimized to mitigate the impact of wireless communications on the convergence of the considered wireless FL algorithm. This motivates us to design a proper joint device selection and power control scheme.
\end{remark}

\section{Convergence Analysis}\label{sec:convergence}
In this section, we analyze the convergence behavior of the considered wireless FL algorithm presented in Section \ref{sec:system} for the general smooth non-convex learning problems. We first present the assumptions and preliminaries, and then present the theoretical results on convergence for the considered wireless FL algorithm in terms of the optimality gap between the expected and optimal global loss values.
\subsection{Preliminaries}\label{subsec:pre_convergence}
First, we make the following standard assumptions that are commonly adopted in the convergence analysis of the BCA-type FL algorithms \cite{XLi,XCao,XFan,YWang,XLian}.
\begin{assumption}[\textbf{L-smooth}]\label{as:smooth}
	The global loss function is differentiable and the gradient is uniformly Lipschitz continuous with a positive Lipschitz constant $L$, i.e., $\forall \mathbf{v},\mathbf{w}\in\mathbb{R}^d$,
	\begin{equation}\label{eq:smooth}
		\|\nabla F(\mathbf{v})-\nabla F(\mathbf{w})\|_2\leq L\|\mathbf{v}-\mathbf{w}\|_2,
	\end{equation}
	which is equivalent to 
	\begin{equation}\label{eq:smooth_1}
		F(\mathbf{v})\leq F(\mathbf{w})+(\mathbf{v}-\mathbf{w})^T\nabla F(\mathbf{w})+\frac{L}{2}\|\mathbf{v}-\mathbf{w}\|_2^2.
	\end{equation}
\end{assumption}
\begin{assumption}[\textbf{Unbiased and bounded variance mini-batch SGD}]\label{as:bounded_sgd}
	The mini-batch SGD is unbiased, i.e., 
	\begin{equation}\label{eq:unbiasedSGD}
		\mathbb{E}[\nabla F_k(\mathbf{w};\mathcal{B})]=\nabla F_k(\mathbf{w}),
	\end{equation}
	and the variance of stochastic gradients in each device is bounded, i.e.,
	\begin{equation}\label{eq:bounded_variance}
		\mathbb{E}[\|\nabla F_k(\mathbf{w};\mathcal{B})-\nabla F_k(\mathbf{w})\|_2^2]\leq\mu^2.
	\end{equation}
\end{assumption}
\begin{assumption}[\textbf{Bounded Gradient Divergence}]\label{as:non_iid}
	The variance of the local gradients for each local device is bounded as
	\begin{equation}\label{eq:non_iid}
		\mathbb{E}\left[\|\nabla F_k(\mathbf{w})-\nabla F(\mathbf{w})\|_2^2\right]\leq\delta,
	\end{equation}
	where $\delta$ measures the data heterogeneity \cite{XZhang}.
\end{assumption}
Based on the global model broadcasting in \eqref{eq:estimate_model}, the mini-batch local SGD \eqref{eq:local_sgd} and \eqref{eq:local_sgd_gradient}, and the proposed adaptive reweighing model aggregation scheme \eqref{eq:propose_aggregation}, the PS updates the global model at the $(t+1)$-th round as 
\begin{equation}
	\begin{aligned}
		&\mathbf{w}(t+1)\\
		=&\sum_{k\in\mathcal{K}}\rho_k(t)\mathbf{w}_k^{E}(t)+\tilde{\mathbf{n}}(t)\\
		\overset{(a)}{=}&\sum_{k\in\mathcal{K}}\rho_k(t)\left[\mathbf{w}_k^{0}(t)-\eta_{t}\sum_{\tau=1}^{E}\nabla F_k(\mathbf{w}_k^{\tau-1}(t);\mathcal{B}_k^{\tau}(t))\right]\\
		&+\tilde{\mathbf{n}}(t)\\
		\overset{(b)}{=}&\sum_{k\in\mathcal{K}}\rho_k(t)\bar{\mathbf{w}}_k(t)+\tilde{\mathbf{n}}(t)\\
		&-\eta_{t}\sum_{k=1}^K\rho_k(t)\sum_{\tau=1}^{E}\nabla F_k(\mathbf{w}_k^{\tau-1}(t);\mathcal{B}_k^{\tau}(t))\\
		\overset{(c)}{=}&\mathbf{w}(t)+\sum_{k\in\mathcal{K}}\frac{\rho_k(t)}{p_s(t)h_k^{\rm dl}(t)}\mathbf{n}_k(t)+\tilde{\mathbf{n}}(t)\\
		&-\eta_{t}\sum_{k=1}^K\rho_k(t)\sum_{\tau=1}^{E}\nabla F_k(\mathbf{w}_k^{\tau-1}(t);\mathcal{B}_k^{\tau}(t)),
	\end{aligned}
\end{equation}
where $(a)$ follows the local mini-batch SGD iterations in \eqref{eq:local_sgd}, $(b)$ is to due to $\mathbf{w}_k^{0}(t)=\bar{\mathbf{w}}_k(t)$, and $(c)$ follows \eqref{eq:estimate_model} and $\sum_{k\in\mathcal{K}}\rho_k(t)=1$.

Define the local model difference $\Delta\mathbf{w}_k(t)$ as
\begin{equation}\label{eq:local_model_diff}
	\begin{aligned}
		\Delta\mathbf{w}_k(t)&=\mathbf{w}_k^{E}(t)-\mathbf{w}_k^{0}(t)\\
		&=-\eta_{t}\sum_{\tau=1}^{E}\nabla F_k(\mathbf{w}_k^{\tau-1}(t);\mathcal{B}_k^{\tau}(t)),
	\end{aligned}
\end{equation}
and define the virtual noisy global model $\tilde{\mathbf{w}}(t)$ as
\begin{equation}\label{eq:noisy_global_model}
	\tilde{\mathbf{w}}(t)=\mathbf{w}(t)+\sum_{k\in\mathcal{K}}\frac{\rho_k(t)}{p_s(t)h_k^{\rm dl}(t)}\mathbf{n}_k(t).
\end{equation}

Thus, we can further rewrite $\mathbf{w}(t+1)$ as
\begin{equation}\label{eq:short_global_aggre}
	\mathbf{w}(t+1)=\tilde{\mathbf{w}}(t)+\sum_{k=1}^K\rho_k(t)\Delta\mathbf{w}_k(t)+\tilde{\mathbf{n}}(t).
\end{equation}

\subsection{Theoretical Results on Convergence}
We first present the following lemmas and their proofs that are used in the proof of Theorem \ref{th:opt_gap}.
\begin{lemma}\label{lem:unbias_bounded_NoisyGlobalModel}
	The virtual noisy global model $\tilde{\mathbf{w}}(t)$ defined in \eqref{eq:noisy_global_model} is unbiased, i.e.,
	\begin{equation}\label{eq:lem5_unbias}
		\mathbb{E}\left[\tilde{\mathbf{w}}(t)\right]=\mathbf{w}(t),
	\end{equation}
	and the variance is bounded as
	\begin{equation}\label{eq:lem5_bounded}
		\mathbb{E}\left[\|\tilde{\mathbf{w}}(t)-\mathbf{w}(t)\|_2^2\right]=\frac{d\sigma_d^2}{p_s^2(t)}\sum_{k\in\mathcal{K}}\frac{\rho_k^2(t)}{|h_k^{\rm dl}(t)|^2}.
	\end{equation}
\end{lemma}

\begin{IEEEproof}
	According to \eqref{eq:noisy_global_model}, we have $\mathbb{E}[\tilde{\mathbf{w}}(t)-\mathbf{w}(t)]=\sum_{k\in\mathcal{K}}\frac{\rho_k(t)}{p_s(t)h_k^{\rm dl}(t)}\mathbb{E}[\mathbf{n}_k(t)]=0$, since $\mathbf{n}_k(t)$'s are i.i.d. following the distribution $\mathcal{CN}(\mathbf{0},\sigma_d^2\mathbf{I})$, which proves \eqref{eq:lem5_unbias}. Similarly, we have 
	\begin{equation}\label{eq:noisy_global_model_bound}
		\begin{aligned}
			\mathbb{E}\left[\|\tilde{\mathbf{w}}(t)-\mathbf{w}(t)\|_2^2\right]&=\mathbb{E}\left[\left\|\sum_{k\in\mathcal{K}}\frac{\rho_k(t)}{p_s(t)h_k^{\rm dl}(t)}\mathbf{n}_k(t)\right\|_2^2\right]\\
			&=\sum_{k\in\mathcal{K}}\left|\frac{\rho_k(t)}{p_s(t)h_k^{\rm dl}(t)}\right|^2\mathbb{E}[\|\mathbf{n}_k(t)\|_2^2]\\
			&=\frac{d\sigma_d^2}{p_s^2(t)}\sum_{k\in\mathcal{K}}\frac{\rho_k^2(t)}{|h_k^{\rm dl}(t)|^2},
		\end{aligned}
	\end{equation}
	which proves \eqref{eq:lem5_bounded}.
\end{IEEEproof}

\begin{lemma}\label{lem:bounded_model_difference}
	The expectation of the square norm of the model difference at each round for device $k$ is bounded by
	\begin{equation}\label{eq:lem2}
		\begin{aligned}
			&\mathbb{E}\left[\|\Delta\mathbf{w}_k(t)\|_2^2\right]\\
			\leq& \eta_{t}^2E^2\left(\mu^2+4\delta\right)+2E\eta_{t}^2L^2\sum_{\tau=1}^{E}\mathbb{E}\left[\|\mathbf{w}_k^{\tau-1}(t)-\mathbf{w}(t)\|_2^2\right]\\
			&+4\eta_{t}^2E^2\mathbb{E}\left[\|\nabla F(\mathbf{w}(t))\|_2^2\right].
		\end{aligned}
	\end{equation}
\end{lemma}
\begin{lemma}\label{lem:bounded_SGD_iteration}
	The sum of the expected square norm of the difference between the local updated model at each SGD iteration and the previous global model is bounded by
	\begin{equation}\label{eq:lem3}
		\begin{aligned}
			&\sum_{\tau=1}^{E}\mathbb{E}\left[\|\mathbf{w}_k^{\tau-1}(t)-\mathbf{w}(t)\|_2^2\right]\\
			\leq&\frac{\frac{2d\sigma_d^2E}{p_s^2(t)}\sum_{k\in\mathcal{K}}\frac{\rho_k^2(t)}{|h_k^{\rm dl}(t)|^2}+2\eta_{t}^2E^3\left(\mu^2+4\delta\right)}{1-4\eta_{t}^2E^2L^2}\\
			&+\frac{+8\eta_{t}^2E^3\mathbb{E}\left[\|\nabla F(\mathbf{w}(t))\|_2^2\right]}{1-4\eta_{t}^2E^2L^2}.
		\end{aligned}
	\end{equation}
\end{lemma}
\begin{lemma}\label{lem:bounded_global_gradient}
	The expectation of the square norm of the gradient of the global loss function at each round is bounded by
	\begin{equation}\label{eq:lem4}
		\|\nabla F(\mathbf{w}(t))\|_2^2\leq 2L\left(F(\mathbf{w}(t))-F^*\right),
	\end{equation}
	where $F^*$ is the optimal global loss function value.
\end{lemma}
The proofs of Lemma \ref{lem:bounded_model_difference}, \ref{lem:bounded_SGD_iteration}, \ref{lem:bounded_global_gradient} follow the same ideas in \cite{YWang}, \cite{XLian}, \cite{YJCho}, with a slight modification according to the problem that we consider. Now, we present the main convergence analysis results in the following theorem.
\begin{theorem}\label{th:opt_gap}
	Let Assumption \eqref{as:smooth}-\eqref{as:non_iid} be hold. Assume the FL algorithm terminates after $T$ rounds, given an initial global model $\mathbf{w}(1)$, the expected optimality gap between the expected and optimal global loss values $\mathbb{E}[F(\mathbf{w}(T+1))]-F^*$ is bounded by
	\begin{align}\label{eq:th1}
			&\mathbb{E}[F(\mathbf{w}(T+1))]-F^*\nonumber\\
			\leq&\prod_{t=1}^TA(t)\mathbb{E}[F(\mathbf{w}(1))-F^*]+\sum_{t=1}^{T-1}\left(\prod_{i=t+1}^{T}A(i)\right)G(t)\nonumber\\
			&+G(T),
	\end{align}
	with
	\begin{equation}\label{eq:A_t}
		A(t)=1+\frac{\eta_{t}EL(20\eta_{t}^2E^2L^2+16\eta_{t}EL-1)}{1-4\eta_{t}^2E^2L^2},
	\end{equation}
	and
	\begin{small}
		\begin{equation}\label{eq:G_t}
			\begin{aligned}
				G(t)=&\underbrace{\left(\frac{2\eta_{t}^2E^2L(1+\eta_tEL)}{1-4\eta_{t}^2E^2L^2}\right)\left(\mu^2+4\delta\right)}_{(a)}\\
				&+\underbrace{\eta_{t}\delta E\left(\sum_{k=1}^K\frac{1}{q_k}+1\right)}_{(b)}\nonumber\\
				&+\underbrace{\left(\frac{d\sigma_d^2L(1+2\eta_tE+4\eta_t^2E^2L^2)}{(1-4\eta_{t}^2E^2L^2)p_s^2(t)}\right)\sum_{k\in\mathcal{K}}\frac{\rho_k^2(t)}{|h_k^{\rm dl}(t)|^2}}_{(c)}\\
				&+\underbrace{\frac{2d\sigma_u^2L}{\left(\sum_{j\in\mathcal{K}}a_j(t)p_j(t)\right)^2}}_{(d)}.
			\end{aligned}
		\end{equation}
	\end{small}
\end{theorem}
\begin{IEEEproof}
	Please see Appendix. \ref{appen:th1_proof}.
\end{IEEEproof}
\begin{remark}\label{rem:convergence}
	The expected optimality gap between the expected and optimal global loss values given in the right hand side (RHS) of \eqref{eq:th1} reveals several important insights: 
	\begin{enumerate}
		\item When the learning rate is set small enough, i.e., $\eta_{t}\leq\frac{1}{20EL}$, we have $A(t)<1$, which implies $\lim_{T\rightarrow\infty}\prod_{t=1}^TA(t)=0$. In this case, the proposed wireless FL algorithm converges to a biased solution, and the expected optimality gap $\mathbb{E}[F(\mathbf{w}(T+1))]-F^*$ is upper bounded only by the the last two terms on the RHS of \eqref{eq:th1}, which is a linear combination of the performance gap $G(t)$ in each round given in \eqref{eq:G_t}.
		\item The performance gap $G(t)$ in each round consists of 4 terms $(a)$-$(d)$ with clear physical meanings: term $(a)$ is caused by gradient variance and data heterogeneity, term $(b)$ is caused solely by data heterogeneity, and terms $(c)$ and $(d)$ are caused by the downlink and uplink wireless communications, respectively.
		\item If the FL algorithm is performed in the ideal environment, where the broadcasting and model aggregation are not decayed by wireless channels, including the channel fading and additive noise, terms $(c)$ and $(d)$ equals to zero. In this case, the proposed FL algorithm still converges to a biased solution due to the impact of gradient variance and data heterogeneity, which coincides the analysis of inconsistency issue of FL in \cite{JWang}. 
	\end{enumerate}
\end{remark}

\section{Optimality Gap Minimization}\label{sec:gap_min}
In this section, we present the joint device selection and power control to minimize the optimimality gap between the expected and optimal global loss values derived in Theorem \ref{th:opt_gap} for the considered wireless FL system under both the individual and sum power constraints, respectively.
\subsection{Problem Formulation}
As mentioned in Remark \ref{rem:convergence}, by properly choosing the learning rate $\eta_{t}\leq\frac{1}{20EL}$ to guarantee the convergence of the proposed wireless FL algorithm, the optimality gap between the expected and optimal global loss values derived in \eqref{eq:th1} becomes
\begin{equation}\label{eq:opt_gap}
	\Lambda = \sum_{t=1}^{T-1}\left(\prod_{i=t+1}^{T}A(i)\right)G(t)+G(T),
\end{equation}
which determines the performance of the algorithm when converges and needs to be minimized. Specifically, with instantaneous CSI in each round, we minimize the performance gap $G(t)$ over the device selection indicator $a_k(t)$, downlink transmit power value $p_s(t)$ and uplink transmit power value $p_k(t)$ round by round. Besides, we ignore the constant terms $(a)$ and $(b)$ in $G(t)$ that is related to the gradient variance and data heterogeneity, and focus on minimizing the sum of terms $(c)$ and $(d)$ in $G(t)$, since the goal of this paper is to minimize the impact of the downlink and uplink wireless communications on the convergence of the considered wireless FL algorithm. Hence, by dropping the notation  $t$ for simplicity, we can formulate the following performance gap minimization problem as
\begin{subequations}\label{opt:problem}
	\begin{align} 			   
		\underset{\{\mathbf{a}, p_s,\mathbf{p}\}}{\text{min}}\ \ &\left(\frac{d\sigma_d^2L(1+2\eta E+4\eta^2E^2L^2)}{(1-4\eta^2E^2L^2)p_s^2}\right)\sum_{k\in\mathcal{K}}\frac{\rho_k^2}{|h_k^{\rm dl}|^2}\nonumber\\
		&+\frac{2d\sigma_u^2L}{\left(\sum_{j\in\mathcal{K}}a_jp_j\right)^2}\label{opt:obj}\\
		{\text{s. t.}}\ \ &a_k\in\{0,1\},\label{opt:a_k}\\
		&p_s^2\leq \bar{P}_{\rm dl},\label{opt:Ps_max}\\ 
		&\mathbf{p}\in\mathcal{P},\label{opt:Pk_max}
	\end{align}
\end{subequations} 
where $\mathbf{a}=[a_1,\cdots,a_K]$, $\mathbf{p}=[p_1,\cdots,p_k]$, \eqref{opt:a_k} denotes the feasible set of the selection indicators, and \eqref{opt:Ps_max} is the downlink transmit power constraint with $\bar{P}_{\rm dl}=P_{\rm max}^{\rm dl}/\|\mathbf{w}\|_2^2$.
For the individual power constraint, $\mathcal{P}$ in \eqref{opt:Pk_max} is given as 
\begin{align}
	\mathcal{P}=\left\{\mathbf{p}:\mathbf{a}^T\mathbf{p}^T\mathbf{Q}_k\mathbf{p}\mathbf{a}-P_{\rm max}^k\leq 0,k=1,\cdots,K\right\},
\end{align}
which is the matrix-vector form of \eqref{eq:individual_power_constraint}, with $\mathbf{Q}_k\in\mathbb{R}^{K\times K}, k=1,\cdots,K$, being a diagonal matrix with $[\mathbf{Q}_k]_{k,k}=\frac{\|\mathbf{w}_k^E\|_2^2}{|h_k^{\rm up}|^2}$ and all other entries being zero; for the sum power constraint, $\mathcal{P}$ in \eqref{opt:Pk_max} is given as
\begin{align}
	\mathcal{P}=\left\{\mathbf{p}:\mathbf{a}^T\mathbf{p}^T\mathbf{Q}_0\mathbf{p}\mathbf{a}-P_{\rm tot}\leq 0\right\},
\end{align}
which is the matrix-vector form of \eqref{eq:sum_power_constraint}, with $\mathbf{Q}_0=\mathtt{diag}\left(\frac{\|\mathbf{w}_1^E\|_2^2}{|h_1^{\rm up}|^2},\cdots,\frac{\|\mathbf{w}_K^E\|_2^2}{|h_K^{\rm up}|^2}\right)\in\mathbb{R}^{K\times K}$.
\begin{remark}
	Intuitively, directly minimizing $\Lambda$ in \eqref{eq:opt_gap} over multiple rounds with non-causally known CSI could result in better performance than minimizing $G(t)$ round by round. However, the unavailability of non-causal information of model parameter $\mathbf{w}(t)$ and $\mathbf{w}_k^E(t)$ at the PS and device $k$, which determines the downlink and uplink transmit power constraints \eqref{opt:Ps_max} and \eqref{opt:Pk_max}, hinders us to directly apply joint optimization over multiple rounds. Though some existing literature introduces additional assumption that $\mathbb{E}\|\mathbf{w}(t)\|_2^2$ or $\mathbb{E}\|\mathbf{w}_k^E(t)\|_2^2$ is upper bounded by some constants \cite{SWang, XCao2}, which can avoid the requirement of non-causal model information, such assumption is not satisfied for many practical situations, especially when the model dimension is large, and is thus beyond scope of this paper. Hence,  this paper chooses to minimize $G(t)$ round by round with only causal CSI, since $\Lambda$ is a linear combination of $G(t)$, as discussed above. 
\end{remark}

\subsection{Optimal Solution}\label{subsec:solution}
In this section, we propose a joint device selection and power control algorithm to solve problem \eqref{opt:problem}. Obviously, the formulated optimization problem \eqref{opt:problem} is a typical MIP problem, where the difficulty for solving this problem is the binary selection variables $\mathbf{a}$ and the non-convex objective function \eqref{opt:obj}. 

However, we could replace the device selection $\mathbf{a}$ with the uplink transmit power value $\mathbf{p}$, based on the observation that the selection indicator $\mathbf{a}$ is always coupled with uplink transmit power value $\mathbf{p}$, and independent of the downlink transmit power value $p_s$, as shown in \eqref{opt:problem}. Hence, the device is selected only when its uplink transmit power value $p_k$ is positive. Thus,  we can drop the selection indicator $\mathbf{a}$ in \eqref{opt:obj} and \eqref{opt:Pk_max} and its associated constraint \eqref{opt:a_k}. Besides, to minimize the objective function \eqref{opt:obj}, the constraint \eqref{opt:Ps_max} should be met with equality, i.e., $p_s^2=\bar{P}_{\rm dl}$, since \eqref{opt:obj} monotonically decreases with $p_s^2$. Hence, problem \eqref{opt:problem} can be equivalently transformed as
\begin{subequations}\label{opt2:problem}	
	\begin{align}	   
		\underset{\mathbf{p}}{\text{min}}\ \ &\left(\frac{d\sigma_d^2L(1+2\eta E+4\eta^2E^2L^2)}{(1-4\eta^2E^2L^2)\bar{P}_{\rm dl}}\right)\sum_{k\in\mathcal{K}}\frac{\rho_k^2}{|h_k^{\rm dl}|^2}\nonumber\\
		&+\frac{2d\sigma_u^2L}{\left(\sum_{j\in\mathcal{K}}p_j\right)^2}\label{opt2:obj}\\
		{\text{s. t.}}\ \ & \mathbf{p}\in\mathcal{P}_0. \label{opt2:Pk_max} 
	\end{align}
\end{subequations}
For the individual uplink power constraint case, $\mathcal{P}_0$ is given as 
\begin{align}\label{eq:up_pmax_ind}
	\mathcal{P}_0=\left\{\mathbf{p}:\mathbf{p}^T\mathbf{Q}_k\mathbf{p}-P_{\rm max}^k\leq 0,k=1,\cdots,K\right\},
\end{align}
and for the sum uplink power constraint case, $\mathcal{P}_0$ is given as 
\begin{align}\label{eq:up_pmax_sum}
	\mathcal{P}_0=\left\{\mathbf{p}:\mathbf{p}^T\mathbf{Q}_0\mathbf{p}-P_{\rm tot}\leq 0\right\},
\end{align}

By rearranging the objective function \eqref{opt2:obj} in the matrix-vector form, problem \eqref{opt2:problem} is equivalent to 
\begin{equation}\label{opt3:problem}		
	\begin{aligned}   
		\underset{\mathbf{p}}{\text{min}}\ \ &\frac{\mathbf{p}^T\mathbf{\Theta}\mathbf{p}+2d\sigma_u^2L}{\mathbf{p}^T\mathbf{1}\mathbf{1}^T\mathbf{p}}\\
		{\text{s. t.}}\ \ & \eqref{opt2:Pk_max},
	\end{aligned}
\end{equation}
where  $\mathbf{\Theta}=\mathtt{diag}(\mathbf{\theta})\in\mathbb{R}^{K\times K}$ with $\mathbf{\theta}=[\theta_1,\cdots,\theta_K]\in\mathbb{R}^K$ and $\theta_k=\frac{d\sigma_d^2L(1+2\eta E+4\eta^2E^2L^2)}{(1-4\eta^2E^2L^2)\bar{P}_{\rm dl}|h_k^{\rm dl}|^2}$. Now, problem \eqref{opt3:problem} is a typical QCRQ minimization problem. It has been proved in \cite{ABeck} that with the given constraint \eqref{opt2:Pk_max} for both the individual and sum uplink power constraint cases \eqref{eq:up_pmax_ind} and \eqref{eq:up_pmax_sum}, the QCRQ problem \eqref{opt3:problem} can be equivalently rewritten as the following homogeneous version by introducing a real auxiliary variable $s$ with $\mathbf{y}=s\mathbf{p}$, i.e.,
\begin{subequations}\label{opt_homo:problem}
	\begin{align} 			   
		\underset{\{\mathbf{y},s\}}{\text{min}}\ \ &\mathbf{y}^T\mathbf{\Theta}\mathbf{y}+2d\sigma_u^2Ls^2\label{opt_homo:obj}\\
		{\text{s. t.}}\ \ &\mathbf{y}^T\mathbf{1}\mathbf{1}^T\mathbf{y}=1,\label{opt_homo:dem=1}\\
		&\mathbf{y}\in\mathcal{Y},\label{opt_homo:up_power}
	\end{align}
\end{subequations}
where for the individual uplink power constraint case, $\mathcal{Y}$ in \eqref{opt_homo:up_power} is given as
\begin{equation}\label{eq:y_ind}
	\mathcal{Y}=\left\{\mathbf{y}:\mathbf{y}^T\mathbf{Q}_k\mathbf{y}-P_{\rm max}^ks^2\leq 0,k=1,\cdots,K\right\},
\end{equation}
and for the sum uplink power constraint case, $\mathcal{Y}$ in \eqref{opt_homo:up_power} is given as 
\begin{equation}\label{eq:y_sum}
	\mathcal{Y}=\left\{\mathbf{y}:\mathbf{y}^T\mathbf{Q}_0\mathbf{y}-P_{\rm tot}s^2\leq 0\right\},
\end{equation}

Though problem \eqref{opt_homo:problem} is still non-convex, it can be approximated by its SDR \cite{ZQLuo}. First, we make the change of variables $\mathbf{z}=(\mathbf{y}^T,s)^T$, and rewrite \eqref{opt_homo:problem} as 
\begin{subequations}\label{sdr_opt:problem}
	\begin{align} 			   
		\underset{\mathbf{z}}{\text{min}}\ \ &\mathbf{z}^T\tilde{\mathbf{\Theta}}\mathbf{z}\label{sdr_opt:obj}\\
		{\text{s. t.}}\ \ &\mathbf{z}^T\mathbf{C}\mathbf{z}=1,\label{sdr_opt:dem=1}\\
		&\mathbf{z}\in\mathcal{Z},\label{sdr_opt:up_power}
	\end{align}
\end{subequations} 
where $\tilde{\mathbf{\Theta}}$ and $\mathbf{C}$ are given as
\begin{equation}\label{eq:tilde_Phi_C}
	\tilde{\mathbf{\Theta}}=\left(
	\begin{array}{cc}
		\mathbf{\Theta} & \mathbf{0}\\
		\mathbf{0}^T & 2d\sigma_u^2L
	\end{array}
	\right),\ 
	\mathbf{C}=\left(
	\begin{array}{cc}
		\mathbf{1}\mathbf{1}^T & \mathbf{0}\\
		\mathbf{0}^T & 0
	\end{array}
	\right).
\end{equation}
For the individual uplink power constraint case, $\mathcal{Z}$ in \eqref{sdr_opt:up_power} is given as 
\begin{equation}
	\mathcal{Z}=\left\{\mathbf{z}:\mathbf{z}^T\tilde{\mathbf{Q}}_k\mathbf{z}\leq0,k=1,\cdots,K\right\},
\end{equation}
where $\tilde{\mathbf{Q}}_k$ is given as
\begin{equation}
	\tilde{\mathbf{Q}}_k=\left(
	\begin{array}{cc}
		\mathbf{Q}_k & \mathbf{0}\\
		\mathbf{0}^T & -P_{\rm max}^k
	\end{array}
	\right),
\end{equation}
and for the sum uplink power constraint case, $\mathcal{Z}$ in \eqref{sdr_opt:up_power} is given as 
\begin{equation}
	\mathcal{Z}=\left\{\mathbf{z}:\mathbf{z}^T\tilde{\mathbf{Q}}_0\mathbf{z}\leq0\right\},
\end{equation}
where $\tilde{\mathbf{Q}}_0$ is given as
\begin{equation}
	\tilde{\mathbf{Q}}_0=\left(
	\begin{array}{cc}
		\mathbf{Q}_0 & \mathbf{0}\\
		\mathbf{0}^T & -P_{\rm tot}
	\end{array}
	\right).
\end{equation}

Next, by introducing a new variable $\mathbf{Z}=\mathbf{z}\mathbf{z}^T$, which is positive semidefinite (PSD) with rank-one, problem \eqref{sdr_opt:problem} is equivalent to 
\begin{subequations}\label{sdr_opt2:problem}
	\begin{align} 			   
		\underset{\mathbf{Z}}{\text{min}}\ \ &\mathtt{tr}(\tilde{\mathbf{\Theta}}\mathbf{Z})\label{sdr_opt2:obj}\\
		{\text{s. t.}}\ \ &\mathtt{tr}(\mathbf{C}\mathbf{Z})=1,\label{sdr_opt2:dem=1}\\
		&\mathbf{Z}\in\mathcal{Z}_0,\label{sdr_opt2:up_power}\\
		&\mathtt{rank}(\mathbf{Z})=1,\label{sdr_opt2:rank1}\\
		&\mathbf{Z}\succeq 0,\label{sdr_opt2:psd}
	\end{align}
\end{subequations}
where for the individual uplink transmit power constraint case, we obtain
\begin{equation}
	\mathcal{Z}_0=\left\{\mathbf{Z}:\mathtt{tr}(\tilde{\mathbf{Q}}_k\mathbf{Z})\leq 0,k=1,\cdots,K\right\};
\end{equation}
for the sum uplink transmit power constraint case, we obtain
\begin{equation}
	\mathcal{Z}_0=\left\{\mathbf{Z}:\mathtt{tr}(\tilde{\mathbf{Q}}_0\mathbf{Z})\leq 0\right\}.
\end{equation}

Now, problem \eqref{sdr_opt2:problem} is still non-convex due to the non-convex rank-one constrain \eqref{sdr_opt2:rank1}. However, we can ignore the rank-one constraint \eqref{sdr_opt2:rank1} and consider the relaxed version of problem \eqref{sdr_opt2:problem} as 
\begin{subequations}\label{sdr_opt3:problem}	
	\begin{align}	   
		\underset{\mathbf{Z}}{\text{min}}\ \ &\mathtt{tr}(\tilde{\mathbf{\Theta}}\mathbf{Z}),\\
		{\text{s. t.}}\ \ & \eqref{sdr_opt2:dem=1},\eqref{sdr_opt2:up_power}, \eqref{sdr_opt2:psd},
	\end{align}
\end{subequations}
which is convex and can be efficiently solved by using CVX \cite{CVX}, where we denote the optimal solution to problem \eqref{sdr_opt3:problem} as $\mathbf{Z}^*$. In the following Theorem \ref{th:opt_solution}, we show how to directly derive the optimal solution to the original problem \eqref{opt2:problem} based on $\mathbf{Z}^*$ for both the individual and sum uplink transmit power constraint cases.

\begin{theorem}\label{th:opt_solution}
	Given the optimal solution $\mathbf{Z}^*$ to problem \eqref{sdr_opt3:problem} for both the individual and sum uplink transmit power constraint cases, we construct its $K$-th order leading principal submatrix $\mathbf{Z}_K^*$ by deleting its $(K+1)$st row and column, which is rank-one and can be decomposed as $\mathbf{Z}_K^*=\mathbf{b}\mathbf{b}^T$.  Then, the optimal solution to the original problem \eqref{opt2:problem} is given as $\mathbf{p}^*=\mathbf{b}/\sqrt{[\mathbf{Z}^*]_{K+1,K+1}}$, where $[\mathbf{Z}^*]_{K+1,K+1}$ is the $(K+1)$-th diagonal element of $\mathbf{Z}^*$.
\end{theorem}
\begin{IEEEproof}
	The rank-one property of $\mathbf{Z}_K^*$ can be explored by utilizing the strong duality between problem \eqref{sdr_opt3:problem} and its dual problem, and the special structure of $\tilde{\mathbf{\Phi}}$, $\mathbf{C}$, $\tilde{\mathbf{Q}}_k$, and $\tilde{\mathbf{Q}}_0$, similar to the  proof of Theorem 1 in \cite{FJiang}. Then, given the rank-one decomposition of $\mathbf{Z}_K^*$ as  $\mathbf{Z}_K^*=\mathbf{b}\mathbf{b}^T$,  it's easy to verify that there always exists a rank-one decomposition of $\mathbf{Z}^*$ as $\mathbf{Z}^*=\tilde{\mathbf{b}}\tilde{\mathbf{b}}^T$ with $\tilde{\mathbf{b}}=\left[\mathbf{b}^T,\sqrt{[\mathbf{Z}^*]_{K+1,K+1}}\right]^T$. Finally, according to Theorem 3.2 in \cite{ABeck}, the optimal solution to the original problem \eqref{opt2:problem} is obtained as $\mathbf{p}^*=\mathbf{b}/\sqrt{[\mathbf{Z}^*]_{K+1,K+1}}$.
\end{IEEEproof}

\begin{algorithm}[!htp]
	\caption{Proposed Wireless FL Algorithm}
	\label{algo}
	\begin{algorithmic}[1]  
		\Require initial global model parameter $\mathbf{w}(1)$; batch size $B$; number of local SGD iterations $E$; learning rate $\eta_{t}$; downlink transmit power budget $P_{\rm max}^{\rm dl}(t)$; individual uplink transmit power budget $P_{\max}^k(t)$; sum uplink transmit power budget $P_{\rm tot}(t)$; communication round budget $T$;
		\For{$t=1:T$} 
		\State Obtain downlink channel gains $\{h_k^{\rm dl}(t)\}_{k=1}^K$ from the PS to terminal devices;
		\State PS broadcasts $\mathbf{w}(t)$ to the terminal devices;
		\For{$k=1:K$}
		\State Terminal device $k$ sets its current model as $\bar{\mathbf{w}}_k(t)$ based on \eqref{eq:rx_model};
		\State Terminal device $k$ updates $\bar{\mathbf{w}}_k(t)$ to $\mathbf{w}_k^E(t)$ via mini-batch SGD \eqref{eq:local_sgd};
		\EndFor
		\State Obtain uplink channel gains $\{h_k^{\rm up}(t)\}_{k=1}^K$ from terminal devices to the PS;
		\State Solve problem \eqref{opt3:problem} via SDR as in Section \ref{sec:gap_min} to obtain optimal $\mathbf{p}^*(t)=[p_1^*(t),\cdots,p_K^*(t)]$;
		\For{$k=1:K$}
		\If{$p_k^*>0$}
		\State Set $a_k(t)=1$ and select device $k$ to upload its updated model $\mathbf{w}_k^E(t)$ with pre-processing factor \eqref{eq:processing_factor};
		\Else
		\State Set $a_k(t)=0$ and device $k$ keeps silent;
		\EndIf
		\EndFor
		\State The PS aggregates the received models and updates the global model as $\mathbf{w}(t+1)$ based on \eqref{eq:propose_aggregation};
		\EndFor
	\end{algorithmic}
\end{algorithm}

Finally, the proposed wireless FL algorithm is summarized in Algorithm \ref{algo}. During each communication round, our proposed device selection and power control scheme only requires to solve one semidefinite programming (SDP) problem in \eqref{sdr_opt3:problem}. Since the number of the constraints is less than the problem size $K+1$, the computation complexity of solving problem \eqref{sdr_opt3:problem} is $\mathcal{O}((K+1)^6)$ \cite{ZQLuo,CHelmberg}, which is polynomial w.r.t. the number of the devices.

\section{Numerical Results}\label{sec:numerical}
This section evaluates the performance of our proposed wireless FL algorithm for image classification on the well-known MNIST \cite{LeCun} and CIFAR-10 \cite{Cifar_dataset} datasets, respectively, where MNIST dataset consists of 10 classes of black-and-white handwriting digits ranging from “0” to “9” with 60000 training and 10000 test samples, and CIFAR-10 dataset consists of 10 classes of colored objects with 50000 training and 10000 test samples.

Here, both the uplink and downlink channels follow the i.i.d. Rayleigh fading, i.e., $h_k^{\rm dl}(t)$ and $h_k^{\rm up}(t),$ are modeled as i.i.d., CSCG random variables with zero mean and unit variance. The downlink and upink noise variances are set as $\sigma_d=\sigma_u=0.1$. Given the model dimension $d$, the downlink transmit power budget $P_{\rm max}^{\rm dl}(t)$ is set as $P_{\rm max}^{\rm dl}(t)=10\times d$ and the individual uplink  transmit power budget $P_{\rm max}^k(t)$ is set as $P_{\rm max}^k(t)=5\times d$, $\forall k,t$. We consider $K=20$ local devices, and set $E=5$, $B=100$, and $L=10$ \cite{MMAmiri_1}. The learning rates for both the two tasks are set as $\eta_t=\frac{1}{20EL}$ with a decaying rate of 0.995 for every 30 communication rounds, which satisfies the condition in Remark \ref{rem:convergence}. We consider balanced and unbalanced data size settings, where we set $D_k = 800,\forall k$ for the balanced data size setting, and we set $D_k$ being uniformly distributed in the range of $[500,1000]$ for the unbalanced data size setting.

For the proposed scheme, we only present the results with the individual uplink transmit power constraint, and the cases with the sum uplink transmit power constraint are omitted for conciseness, since their results are the same as the individual uplink transmit constraints cases.  For performance comparison, we consider three baselines: $a)$ the ideal {\it FedAvg} scheme \cite{HBMcMahan}, where the broadcasting and model aggregation are not decayed by the wireless channels including the channel fading and additive noise, with full device participation and fixed weight $q_k$ in \eqref{FL_problem} for device $k$; $b)$ the device selection with MSE threshold scheme, where maximum downlink power transmission and MSE threshold based device selection \cite{KYang} are adopted; $c)$ the truncated channel inversion scheme, where maximum downlink power transmission and truncated channel inversion for the uplink transmission and model aggregation in \cite{MMAmiri_1} and \cite{Sxia} are adopted. 

\subsection{MNIST dataset}\label{subsec:mnist}
For this task, we aim to train a convolution neural network (CNN) as the classifier model, which consists of three $3\times3$ convolution layers (each with 8, 16, and 32 channels, respectively), each followed by a $2\times 2$ max pooling layer with stride 2; fully connected layer with 10 units; and finally a softmax output layer. All convolutional layers are also followed by batch normalization layers and mapped by ReLU activation. We consider both the i.i.d. and non-i.i.d. cases for this task. For the i.i.d. case, we randomly assign training samples to each device $k$; while for the non-i.i.d. case, we split the training samples into 5 disjoint subsets with each subset containing 2 classes of images, and each subset is chose to randomly assign training samples to $K/5$ devices. Hence, each device only contains two classes of images and different devices contains different classes of images.

\begin{figure}[htbp]
	\centering
	\subfigure[i.i.d. case]{
		\includegraphics[width=0.47\textwidth]{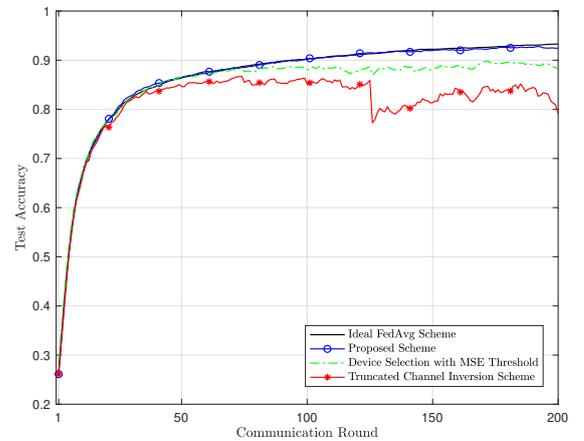}\label{subfig:mnist_iid} 
	}
	\subfigure[non-i.i.d. case]{
		\includegraphics[width=0.47\textwidth]{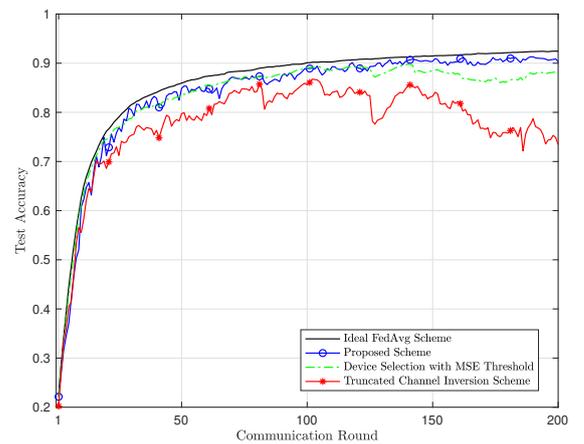} \label{subfig:mnist_noniid} 
	}
	\caption{Test accuracy vs. the communication round with balanced data size.}
	\label{fig:mnist} 
\end{figure}

Fig. \ref{fig:mnist} compares the performance of all the four schemes w.r.t. the communication round in terms of the test accuracy with balanced data size on the MNIST dataset. For the i.i.d. case in Fig. \ref{subfig:mnist_iid}, it is observed that the test accuracies of the ideal {\it FedAvg} scheme, the proposed scheme, and the device selection with MSE threshould scheme increase with the communication round and finally converge. The proposed scheme achieves comparable test accuracy performance to the ideal {\it FedAvg} scheme, where the average accuracy over the last 20 communication rounds for the ideal {\it FedAvg} scheme is $93.09\%$ and the average accuracy over the last 20 communication rounds for the proposed scheme is $92.75\%$. Besides, the proposed scheme outperforms the device selection with MSE threshould scheme, whose average accuracy over the last 20 communication rounds is $89.14\%$. It is observed that all the above three schemes outperform the truncated channel inversion scheme, whose average accuracy over the last 20 communication rounds is $83.34\%$. For the truncated channel inversion scheme, the test accuracy first increase with the communication round during round the first 125 rounds, and then becomes unstable during the rest communication rounds, and does not guarantee convergence. A possible explanation for this phenomenon is that the device selection of the truncated channel inversion scheme is only based on the uplink channel conditions, and the equal weight aggregation in this scheme gives more weight for the local model that damaged by the downlink communications, which degrades the training performance. 

For the non-i.i.d. case in Fig. \ref{subfig:mnist_noniid}, all the schemes suffer from performance degradation compared to the i.i.d. case in Fig. \ref{subfig:mnist_iid}, except for the ideal {\it FedAvg} scheme, whose average accuracy over the last 20 communication rounds is $92.32\%$. The average accuracy over the last 20 communication rounds for the proposed scheme is $90.87\%$, which is slightly worse than that of the ideal {\it FedAvg} scheme, but still larger than that of the device selection with MSE threshold scheme, whose average accuracy over the last 20 communication rounds is $77.86\%$. The truncated channel inversion scheme still has the worst performance, whose convergence is still not guaranteed. Besides, the increasing trend of the proposed scheme becomes less stable compared to that in the i.i.d. case in Fig. \ref{subfig:mnist_iid}, which is obviously caused by the non-i.i.d. data distributions.

\begin{figure}[htbp]
	\centering
	\subfigure[i.i.d. case]{
		\includegraphics[width=0.47\textwidth]{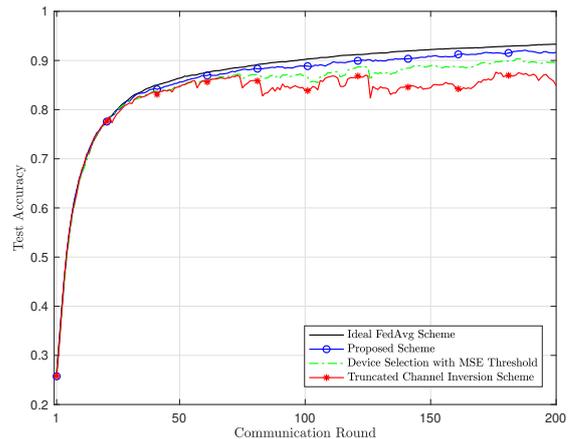}\label{subfig:mnist_iid_un} 
	}
	\subfigure[non-i.i.d. case]{
		\includegraphics[width=0.47\textwidth]{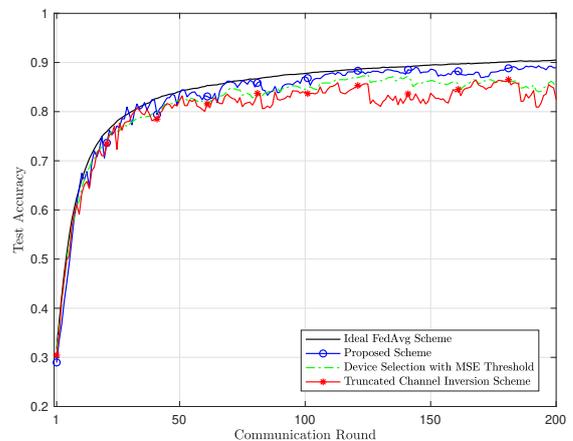} \label{subfig:mnist_noniid_un} 
	}
	\caption{Test accuracy vs. the communication round with unbalanced data size.}
	\label{fig:mnist_un} 
\end{figure}

Fig. \ref{fig:mnist_un} compares the performance of the four schemes w.r.t. the communication round in terms of the test accuracy with unbalanced data size on the MNIST dataset. For the i.i.d. case in Fig. \ref{subfig:mnist_iid_un}, similar to the balanced data size scenario in Fig. \ref{subfig:mnist_iid}, the test accuracies of the ideal {\it FedAvg} scheme, the proposed scheme, and the device selection with MSE threshold scheme increase with the communication round and finally converge. The average accuracy over the last 20 communication rounds of the proposed scheme is $91.73\%$, which is slightly lower than the test accuracy of the ideal {\it FedAvg} scheme, whose average accuracy over the last 20 communication rounds is $93.11\%$, but still larger than both the device selection with MSE threshold and truncated channel inversion schemes and, whose average accuracies over the last 20 communication rounds are $89.74\%$ and $86.77\%$, respectively. 

For the non-i.i.d. case in Fig. \ref{subfig:mnist_noniid_un}, all the four schemes suffer from performance degradation compared to the i.i.d. case in Fig. \ref{subfig:mnist_iid_un}, where the average accuracies over the last 20 communication rounds for the ideal {\it FedAvg}, proposed, device selection with MSE threshold, and truncated channel inversion schemes are $90.3\%$, $88.97\%$, $85.34\%$, and $83.91\%$, respectively. Obviously, the performance of the proposed scheme is still slightly worse than the ideal {\it FedAvg} scheme. Note that even the increasing trends of the test accuracy for the truncated channel inversion scheme are still unstable for both the i.i.d. and non-i.i.d. cases, they shows the convergence trend, which is slightly different from the balanced data size case in Fig. \ref{fig:mnist}. This is possibly due to that the equal weight aggregation for the truncated channel inversion scheme combats the participation of the local model damaged by the downlink communications in certain level, since the weight for perfect model aggregation is inherently different in this unbalanced data size case. However, the performance of our proposed scheme still outperforms the truncated channel inversion scheme with unbalanced data size.

\subsection{CIFAR-10 dataset}\label{subsec:cifar}
For this more challenging task, we aim to train a more complex CNN model, which consists of three $5\times5$ convolution layers (each with 32, 32, and 64 channels, respectively), each followed by a $3\times 3$ max pooling layer with stride 2; two fully connected layers (the first with 64 units and mapped by ReLU activation, and the second with 10 units); and final softmax output layer. All convolutional layers are also followed by batch normalization layers and mapped by ReLU activation. We consider the similar i.i.d. and non-i.i.d. cases to that in Section \ref{subsec:mnist}, where we assign training samples in the same way.

\begin{figure}[htbp]
	\centering
	\subfigure[i.i.d. case]{
		\includegraphics[width=0.47\textwidth]{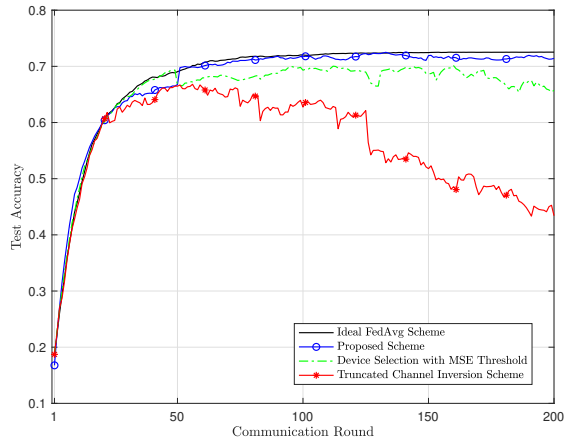}\label{subfig:cifar_iid} 
	}
	\subfigure[non-i.i.d. case]{
		\includegraphics[width=0.47\textwidth]{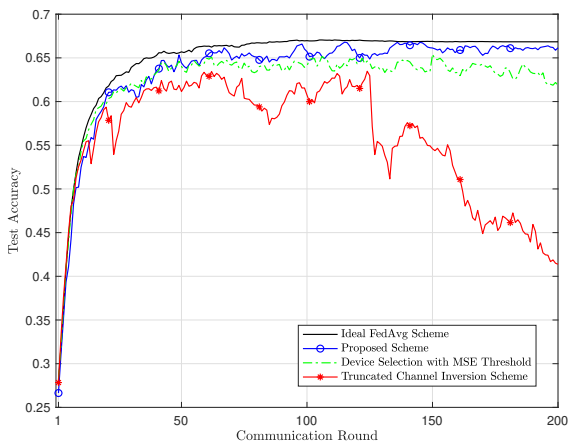} \label{subfig:cifar_noniid} 
	}
	\caption{Test accuracy vs. the communication round with balanced data size.}
	\label{fig:cifar} 
\end{figure}

Fig. \ref{fig:cifar} compares the performance of all the four schemes w.r.t. the communication round in terms of the test accuracy with balanced data size on the CIFAR-10 dataset. For the i.i.d. case in Fig. \ref{subfig:cifar_iid}, it is observed that the test accuracies of the ideal {\it FedAvg} scheme, the proposed scheme, and the device selection with MSE threshold scheme increase with the communication round and finally converge, while the truncated channel inversion scheme fails to converge. The proposed scheme still achieves comparable test accuracy performance to the ideal {\it FedAvg} scheme, where the average accuracy over the last 20 communication rounds for the ideal {\it FedAvg} scheme is $72.52\%$ and the average accuracy over the last 20 communication rounds for the proposed scheme is $71.52\%$, and outperforms the performance of the device selection with MSE threshold scheme, whose average accuracy over the last 20 communication rounds is $66.47\%$. For the non-i.i.d. case in Fig. \ref{subfig:cifar_noniid}, the convergences of the ideal {\it FedAvg} scheme, the proposed scheme, and the device selection with MSE threshold scheme are still guaranteed. However, all the above three schemes suffer from performance degradation compared to the i.i.d. case in Fig. \ref{subfig:mnist_iid}, and their average accuracies over the last 20 communication rounds drop to $66.94\%$, $65.87\%$, and $62.97\%$, respectively. In this case, the truncated channel inversion scheme still fails to converge. 

\begin{figure}[htbp]
	\centering
	\subfigure[i.i.d. case]{
		\includegraphics[width=0.47\textwidth]{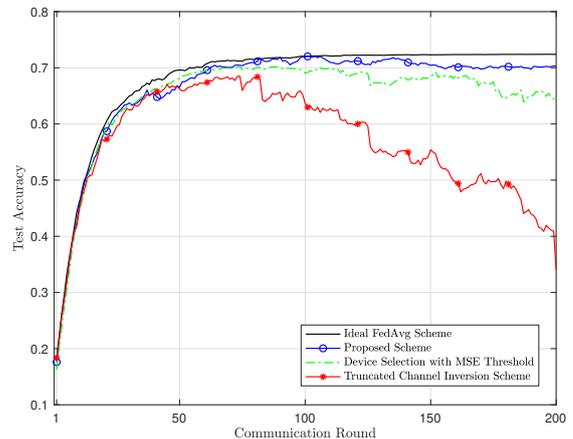}\label{subfig:cifar_iid_un} 
	}
	\subfigure[non-i.i.d. case]{
		\includegraphics[width=0.47\textwidth]{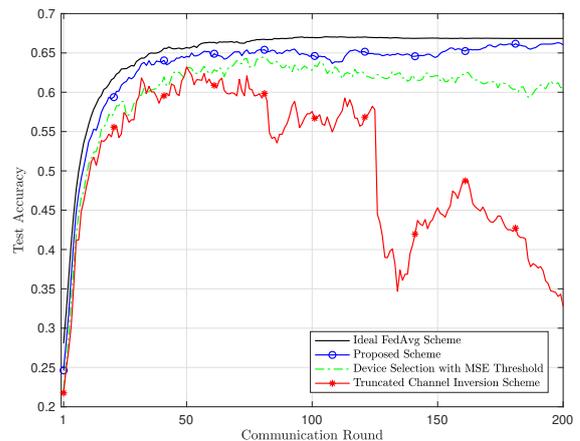} \label{subfig:cifar_noniid_un} 
	}
	\caption{Test accuracy vs. the communication round with unbalanced data size.}
	\label{fig:cifar_un} 
\end{figure}

Fig. \ref{fig:cifar_un} compares the performance of the four schemes w.r.t. the communication round in terms of the test accuracy with unbalanced data size on the CIFAR-10 dataset. For the i.i.d. case in Fig. \ref{subfig:cifar_iid_un}, similar to the balanced data size scenario in Fig. \ref{subfig:cifar_iid}, the test accuracies of the ideal {\it FedAvg} scheme, the proposed scheme, and the device selection with MSE threshold scheme still increase with the communication round and finally converge, while the truncated channel inversion scheme still fails to converge. The performance of the proposed scheme is still slightly worse than the ideal {\it FedAvg} scheme, where the average accuracies over the last 20 communication rounds for the ideal {\it FedAvg} and proposed schemes are $72.39\%$ and $70.14\%$, respectively. The proposed scheme still outperforms the device selection with MSE threshold scheme, whose average accuracies over the last 20 communication rounds is $65.05\%$. For the non-i.i.d. case in Fig. \ref{subfig:cifar_noniid_un}, the truncated channel inversion scheme still fails to converge, while the convergences of the rest three schemes are still be guaranteed. Similar to the balanced data size case in Fig. \ref{fig:cifar}, those three schemes suffer from performance degradation compared to the i.i.d. case in Fig. \ref{subfig:cifar_iid_un}, where the average accuracies over the last 20 communication rounds for the ideal {\it FedAvg} scheme, proposed scheme, and the device selection with MSE threshold scheme drop to $66.84\%$, $65.02\%$, and $60.52\%$, respectively. Obviously, the performance of the proposed scheme is still slightly worse than the ideal {\it FedAvg} scheme, and better than the device selection with MSE threshold scheme. Intuitively, the reason why the truncated channel inversion scheme fails to converge for the CIFAR-10 dataset in Fig. \ref{fig:cifar} and \ref{fig:cifar_un} is the selection criterion is only based on the uplink channel conditions and it hardly able to mitigate the impact of both the downlink and uplink communications on the convergence of FL for a more complex task. 

\section{Conclusion}\label{sec:conclusion}
In this paper, we studied the joint device selection and power control for the wireless FL system, considering both the analog downlink and uplink communications between the PS and terminal devices. We first proposed an AirComp-based adaptive reweighing scheme for model aggregation, where the weights are determined by the selected devices and their uplink transmit power values. Then, we analyzed the convergence behavior of the proposed algorithm, in terms of the expected optimality gap between the expected and optimal global loss values, and revealed how the downlink and uplink wireless communications affects the convergence of the considered FL algorithm. Based on the obtained theoretical results, we formulated an optimality gap minimization problem, for both the individual and sum uplink transmit power constraints, respectively, where we minimized the optimality gap round by round with instantaneous CSI in each round and derived the optimal joint device selection and power control by using the SDR technique for both the two constraint cases. Finally, numerical results showed that the proposed scheme performs very close to the ideal {\it FedAvg} scheme, and significantly outperforms other baseline schemes, i.e., the device selection with MSE threshold and conventional truncated channel inversion AirComp schemes, for both the i.i.d./non-i.i.d data distributions, and balanced/unbalanced data sizes across the devices.

\appendices
\section{Proof of Theorem \ref{th:opt_gap}}\label{appen:th1_proof}
First, according to \eqref{eq:smooth_1}, we have 
\begin{small}
	\begin{align}\label{proof:eq1}
		&F(\mathbf{w}(t+1))\nonumber\\
		\leq& F(\mathbf{w}(t))+\langle\mathbf{w}(t+1)-\mathbf{w}(t),\nabla F\left(\mathbf{w}(t)\right)\rangle\nonumber\\
		&+\frac{L}{2}\|\mathbf{w}(t+1)-\mathbf{w}(t)\|_2^2\nonumber\\
		\overset{(a)}{=}&F(\mathbf{w}(t))+\frac{L}{2}\left\|\tilde{\mathbf{w}}(t)-\mathbf{w}(t)+\sum_{k=1}^K\rho_k(t)\Delta\mathbf{w}_k(t)+\tilde{\mathbf{n}}(t)\right\|_2^2\nonumber\\
		&+\left\langle\tilde{\mathbf{w}}(t)-\mathbf{w}(t)+\sum_{k=1}^K\rho_k(t)\Delta\mathbf{w}_k(t)+\tilde{\mathbf{n}}(t),\nabla F(\mathbf{w}(t))\right\rangle,\nonumber\\
	\end{align}
\end{small}
where $(a)$ follows from \eqref{eq:short_global_aggre}. By taking expectation on both sides of \eqref{proof:eq1}, we obtain 
\begin{small}
	\begin{align}\label{proof:eq2}
		&\mathbb{E}\left[F(\mathbf{w}(t+1))\right]\nonumber\\
		\leq&\mathbb{E}\left[F(\mathbf{w}(t))\right]\nonumber\\
		&+\underbrace{\mathbb{E}\left[\left\langle\tilde{\mathbf{w}}(t)-\mathbf{w}(t)+\sum_{k=1}^K\rho_k(t)\Delta\mathbf{w}_k(t)+\tilde{\mathbf{n}}(t),\nabla F(\mathbf{w}(t))\right\rangle\right]}_{A_1}\nonumber\\
		&+\frac{L}{2}\underbrace{\mathbb{E}\left[\left\|\tilde{\mathbf{w}}(t)-\mathbf{w}(t)+\sum_{k=1}^K\rho_k(t)\Delta\mathbf{w}_k(t)+\tilde{\mathbf{n}}(t)\right\|_2^2\right]}_{A_2}.
	\end{align}
\end{small}
Next, we bound the terms $A_1$ and $A_2$ in the RHS of \eqref{proof:eq2} in the following.

First, we study the term $A_2$.
\begin{small}
	\begin{align}\label{eq:A2_bound}
		&A_2\nonumber\\
		=&\mathbb{E}\left[\left\|\tilde{\mathbf{w}}(t)-\mathbf{w}(t)+\sum_{k=1}^K\rho_k(t)\Delta\mathbf{w}_k(t)+\tilde{\mathbf{n}}(t)\right\|_2^2\right]\nonumber\\
		\overset{(a)}{\leq}&2\mathbb{E}\left[\left\|\tilde{\mathbf{w}}(t)-\mathbf{w}(t)\right\|_2^2\right]+2\mathbb{E}\left[\left\|\sum_{k=1}^K\rho_k(t)\Delta\mathbf{w}_k(t)+\tilde{\mathbf{n}}(t)\right\|_2^2\right]\nonumber\\
		\overset{(b)}{=}&\frac{2d\sigma_d^2}{p_s^2(t)}\sum_{k\in\mathcal{K}}\frac{\rho_k^2(t)}{|h_k^{\rm dl}(t)|^2}+2\mathbb{E}\left[\left\|\sum_{k=1}^K\rho_k(t)\Delta\mathbf{w}_k(t)+\tilde{\mathbf{n}}(t)\right\|_2^2\right]\nonumber\\
		\overset{(c)}{\leq}&\frac{2d\sigma_d^2}{p_s^2(t)}\sum_{k\in\mathcal{K}}\frac{\rho_k^2(t)}{|h_k^{\rm dl}(t)|^2}+4\mathbb{E}\left[\left\|\sum_{k=1}^K\rho_k(t)\Delta\mathbf{w}_k(t)\right\|_2^2\right]\nonumber\\
		&+\frac{4d\sigma_u^2}{\left(\sum_{j\in\mathcal{K}}a_j(t)p_j(t)\right)^2}\nonumber\\
		\overset{(d)}{\leq}&4\sum_{k=1}^K\rho_k(t)\mathbb{E}\left[\|\Delta\mathbf{w}_k(t)\|_2^2\right]+\frac{4d\sigma_u^2}{\left(\sum_{j\in\mathcal{K}}a_j(t)p_j(t)\right)^2}\nonumber\\
		&+\frac{2d\sigma_d^2}{p_s^2(t)}\sum_{k\in\mathcal{K}}\frac{\rho_k^2(t)}{|h_k^{\rm dl}(t)|^2},
	\end{align}
\end{small}
where $(a)$ follows the inequality $\|\mathbf{x}_1+\mathbf{x_2}\|_2^2\leq 2\|\mathbf{x}_1\|_2^2+2\|\mathbf{x}_2\|_2^2$, $(b)$ follows \eqref{eq:lem5_bounded} in Lemma \ref{lem:unbias_bounded_NoisyGlobalModel}, $(c)$ follows the latter inequality again, and uses the fact that $\tilde{\mathbf{n}}(t)=\frac{\mathbf{n}(t)}{\sum_{j\in\mathcal{K}}a_j(t)p_j(t)}$ with $\mathbf{n}(t)$ being the CSCG noise vector, and $(d)$ follows the Jensen's inequality.

Combining \eqref{eq:A2_bound} with Lemmas \ref{lem:bounded_model_difference} and \ref{lem:bounded_SGD_iteration}, we finally bound the term $A_2$ as
\begin{align}\label{eq:A2_bound_final}
	A_2\leq&4\sum_{k=1}^K\rho_k(t)\mathbb{E}\left[\|\Delta\mathbf{w}_k(t)\|_2^2\right]+\frac{4d\sigma_u^2}{\left(\sum_{j\in\mathcal{K}}a_j(t)p_j(t)\right)^2}\nonumber\\
	&+\frac{2d\sigma_d^2}{p_s^2(t)}\sum_{k\in\mathcal{K}}\frac{\rho_k^2(t)}{|h_k^{\rm dl}(t)|^2}\nonumber\\
	\leq&4\eta_{t}^2E^2\left(\mu^2+4\delta\right)+16\eta_{t}^2E^2\mathbb{E}\left[\|\nabla F(\mathbf{w}(t))\|_2^2\right]\nonumber\\
	&+\frac{2d\sigma_d^2}{p_s^2(t)}\sum_{k\in\mathcal{K}}\frac{\rho_k^2(t)}{|h_k^{\rm dl}(t)|^2}+\frac{4d\sigma_u^2}{\left(\sum_{j\in\mathcal{K}}a_j(t)p_j(t)\right)^2}\nonumber\\
	&+8E\eta_{t}^2L^2\sum_{k=1}^K\rho_k(t)\sum_{\tau=1}^{E}\mathbb{E}\left[\|\mathbf{w}_k^{\tau-1}(t)-\mathbf{w}(t)\|_2^2\right]\nonumber\\
	\leq&4\eta_{t}^2E^2\left(\mu^2+4\delta\right)+16\eta_{t}^2E^2\mathbb{E}\left[\|\nabla F(\mathbf{w}(t))\|_2^2\right]\nonumber\\
	&+\frac{2d\sigma_d^2}{p_s^2(t)}\sum_{k\in\mathcal{K}}\frac{\rho_k^2(t)}{|h_k^{\rm dl}(t)|^2}+\frac{4d\sigma_u^2}{\left(\sum_{j\in\mathcal{K}}a_j(t)p_j(t)\right)^2}\nonumber\\
	&+8E\eta_{t}^2L^2\sum_{k=1}^K\rho_k(t)\left(\frac{\frac{2d\sigma_d^2E}{p_s^2(t)}\sum_{k\in\mathcal{K}}\frac{\rho_k^2(t)}{|h_k^{\rm dl}(t)|^2}}{1-4\eta_{t}^2E^2L^2}\right.\nonumber\\
	&+\left.\frac{2\eta_{t}^2E^3\left(\mu^2+4\delta\right)+8\eta_{t}^2E^3\mathbb{E}\left[\|\nabla F(\mathbf{w}(t))\|_2^2\right]}{1-4\eta_{t}^2E^2L^2}\right)\nonumber\\
	=&\frac{4\eta_{t}^2E^2}{1-4\eta_{t}^2E^2L^2}\left(\mu^2+4\delta\right)+\frac{4d\sigma_u^2}{\left(\sum_{j\in\mathcal{K}}a_j(t)p_j(t)\right)^2}\nonumber\\
	&+\frac{16\eta_{t}^2E^2}{1-4\eta_{t}^2E^2L^2}\mathbb{E}\left[\|\nabla F(\mathbf{w}(t))\|_2^2\right]\nonumber\\
	&+\frac{2d\sigma_d^2(1+4\eta_{t}^2E^2L^2)}{(1-4\eta_{t}^2E^2L^2)p_s^2(t)}\sum_{k\in\mathcal{K}}\frac{\rho_k^2(t)}{|h_k^{\rm dl}(t)|^2}
\end{align}

Then, we study the term $A_1$.
\begin{small}
	\begin{align}\label{eq:A1_bound}
		&A_1\nonumber\\
		=&\mathbb{E}\left[\left\langle\tilde{\mathbf{w}}(t)-\mathbf{w}(t)+\sum_{k=1}^K\rho_k(t)\Delta\mathbf{w}_k(t)+\tilde{\mathbf{n}}(t),\nabla F(\mathbf{w}(t))\right\rangle\right]\nonumber\\
		=&\mathbb{E}\left[\left\langle\sum_{k\in\mathcal{K}}\frac{\rho_k(t)}{p_s(t)h_k^{\rm dl}(t)}\mathbf{n}_k(t)+\sum_{k=1}^K\rho_k(t)\Delta\mathbf{w}_k(t)+\tilde{\mathbf{n}}(t),\right.\right.\nonumber\\
		&\left.\left.\nabla F(\mathbf{w}(t))\right\rangle\right]\nonumber\\
		\overset{(a)}{=}&\mathbb{E}\left[\left\langle\sum_{k=1}^K\rho_k(t)\Delta\mathbf{w}_k(t),\nabla F(\mathbf{w}(t))\right\rangle\right]\nonumber\\
		=&-\eta_{t}\sum_{\tau=1}^{E}\mathbb{E}\left[\left\langle\sum_{k=1}^K\rho_k(t)\nabla F_k(\mathbf{w}_k^{\tau-1}(t);\mathcal{B}_k^{\tau}(t)),\nabla F(\mathbf{w}(t))\right\rangle\right]\nonumber\\
		\overset{(b)}{=}&-\eta_{t}\sum_{\tau=1}^{E}\mathbb{E}\left[\left\langle\sum_{k=1}^K\rho_k(t)\nabla F_k(\mathbf{w}_k^{\tau-1}(t)),\nabla F(\mathbf{w}(t))\right\rangle\right]\nonumber\\
		\overset{(c)}{=}&-\frac{\eta_{t}}{2}\sum_{\tau=1}^{E}\mathbb{E}\left[\left\|\sum_{k=1}^K\rho_k(t)\nabla F_k(\mathbf{w}_k^{\tau-1}(t))\right\|_2^2\right]\nonumber\\
		&-\frac{\eta_{t}}{2}\sum_{\tau=1}^{E}\mathbb{E}\left[\|\nabla F(\mathbf{w}(t))\|_2^2\right]\nonumber\\
		&+\frac{\eta_{t}}{2}\sum_{\tau=1}^{E}\mathbb{E}\left[\left\|\sum_{k=1}^K\rho_k(t)\nabla F_k(\mathbf{w}_k^{\tau-1}(t))-\nabla F(\mathbf{w}(t))\right\|_2^2\right]\nonumber\\
		\leq&-\frac{\eta_{t}E}{2}\mathbb{E}\left[\|\nabla F(\mathbf{w}(t))\|_2^2\right]\nonumber\\
		&+\frac{\eta_{t}}{2}\sum_{\tau=1}^{E}\mathbb{E}\left[\left\|\nabla F(\mathbf{w}(t))-\sum_{k=1}^K\rho_k(t)\nabla F_k(\mathbf{w}_k^{\tau-1}(t))\right\|_2^2\right]\nonumber\\
		\overset{(d)}{\leq}&-\frac{\eta_{t}E}{2}\mathbb{E}\left[\|\nabla F(\mathbf{w}(t))\|_2^2\right]\nonumber\\
		&+\underbrace{\eta_{t}\sum_{\tau=1}^{E}\mathbb{E}\left[\left\|\nabla F(\mathbf{w}(t))-\sum_{k=1}^K\rho_k(t)\nabla F_k(\mathbf{w}(t))\right\|_2^2\right]}_{B_1}\nonumber\\
		&+\underbrace{\eta_{t}\sum_{\tau=1}^{E}\mathbb{E}\left[\left\|\sum_{k=1}^K\rho_k(t)\left(\nabla F_k(\mathbf{w}(t))-\nabla F_k(\mathbf{w}_k^{\tau-1}(t))\right)\right\|_2^2\right],}_{B_2}
	\end{align}
\end{small}
where $(a)$ is due to the fact that both $\sum_{k\in\mathcal{K}}\frac{\rho_k(t)}{p_s(t)h_k^{\rm dl}(t)}\mathbf{n}_k(t)$ and $\tilde{\mathbf{n}}(t)$ are zero mean vectors and independent of $\nabla F(\mathbf{w}(t))$,  $(b)$ follows Assumption \ref{as:bounded_sgd}, $(c)$ follows the identity $\langle\mathbf{x}_1,\mathbf{x}_2\rangle=\frac{1}{2}(\|\mathbf{x}_1\|_2^2+\|\mathbf{x}_2\|_2^2-\|\mathbf{x}_1-\mathbf{x}_2\|_2^2)$, and $(d)$ follows the inequality $\|\mathbf{x}_1+\mathbf{x_2}\|_2^2\leq 2\|\mathbf{x}_1\|_2^2+2\|\mathbf{x}_2\|_2^2$.

For \eqref{eq:A1_bound}, we further need to bound the terms $B_1$ and $B_2$. First, we obtain
\begin{small}
	\begin{align}\label{eq:B1_bound}
		B_1=&\eta_{t}\sum_{\tau=1}^{E}\mathbb{E}\left[\left\|\nabla F(\mathbf{w}(t))-\sum_{k=1}^K\rho_k(t)\nabla F_k(\mathbf{w}(t))\right\|_2^2\right]\nonumber\\
		=&\eta_{t}E\mathbb{E}\left[\left\|\sum_{k=1}^Kq_k\nabla F_k(\mathbf{w}(t))-\sum_{k=1}^K\rho_k(t)\nabla F_k(\mathbf{w}(t))\right\|_2^2\right]\nonumber\\
		\overset{(a)}{=}&\eta_{t}E\mathbb{E}\left[\left\|\sum_{k=1}^K(q_k-\rho_k(t))\nabla F_k(\mathbf{w}(t))\right.\right.\nonumber\\
		&\left.\left.-\sum_{k=1}^K(q_k-\rho_k(t))\nabla F(\mathbf{w}(t))\right\|_2^2\right]\nonumber\\
		=&\eta_{t}E\mathbb{E}\left[\left\|\sum_{k=1}^K\frac{q_k-\rho_k(t)}{\sqrt{q_k}}\sqrt{q_k}\left(\nabla F_k(\mathbf{w}(t))\right.\right.\right.\nonumber\\
		&\left.\left.\left.-\nabla F(\mathbf{w}(t))\right)\right\|_2^2\right]\nonumber\\
		\overset{(b)}{\leq}&\eta_{t}E\left(\sum_{k=1}^K\frac{(\rho_k(t)-q_k)^2}{q_k}\right)\sum_{k=1}^Kq_k\mathbb{E}\left[\|\nabla F_k(\mathbf{w}(t))\right.\nonumber\\
		&\left.-\nabla F(\mathbf{w}(t))\|_2^2\right]\nonumber\\
		\overset{(c)}{\leq}&\eta_{t}E\left(\sum_{k=1}^K\frac{1}{q_k}+1\right)\sum_{k=1}^Kq_k\delta\nonumber\\
		=&\eta_{t}\delta E\left(\sum_{k=1}^K\frac{1}{q_k}+1\right),
	\end{align} 
\end{small}
where $(a)$ is due to $\sum_{k=1}^K(q_k-\rho_k(t))=0,\ \forall t$, $(b)$ follows the Cauchy-Schwarz inequality, and $(c)$ follows Assumption \ref{as:non_iid} and $\sum_{k=1}^Kq_k=\sum_{k=1}^K\rho_k(t)=1$.

Then, we bound the term $B_2$ as	
\begin{small}
	\begin{align}\label{eq:B2_bound}
		B_2=&\eta_{t}\sum_{\tau=1}^{E}\mathbb{E}\left[\left\|\sum_{k=1}^K\rho_k(t)\left(\nabla F_k(\mathbf{w}(t))-\nabla F_k(\mathbf{w}_k^{\tau-1}(t))\right)\right\|_2^2\right]\nonumber\\
		\overset{(a)}{\leq}&\eta_{t}\sum_{k=1}^K\rho_k(t)\sum_{\tau=1}^{E}\mathbb{E}\left[\|\nabla F_k(\mathbf{w}(t))-\nabla F_k(\mathbf{w}_k^{\tau-1}(t))\|_2^2\right]\nonumber\\
		\overset{(b)}{\leq}&\eta_{t}L^2\sum_{k=1}^K\rho_k(t)\sum_{\tau=1}^{E}\mathbb{E}\left[\|\mathbf{w}_k^{\tau-1}(t)-\mathbf{w}(t)\|_2^2\right],
	\end{align}
\end{small}
where $(a)$ follows the Jensen's inequality and $(b)$ follows \eqref{eq:smooth} in Assumption \ref{as:smooth}.

Combining \eqref{eq:A1_bound}-\eqref{eq:B2_bound}, we obtain
\begin{align}\label{eq:A1_bound_final}
	A_1\leq&-\frac{\eta_{t}E}{2}\mathbb{E}\left[\|\nabla F(\mathbf{w}(t))\|_2^2\right]+\eta_{t}\delta E\left(\sum_{k=1}^K\frac{1}{q_k}+1\right)\nonumber\\
	&+\eta_{t}L^2\sum_{k=1}^K\rho_k(t)\sum_{\tau=1}^{E}\mathbb{E}\left[\|\mathbf{w}_k^{\tau-1}(t)-\mathbf{w}(t)\|_2^2\right]\nonumber\\
	\overset{(a)}{\leq}&-\frac{\eta_{t}E}{2}\mathbb{E}\left[\|\nabla F(\mathbf{w}(t))\|_2^2\right]+\eta_{t}\delta E\left(\sum_{k=1}^K\frac{1}{q_k}+1\right)\nonumber\\
	&+\eta_{t}L^2\sum_{k=1}^K\rho_k(t)\left(\frac{\frac{2d\sigma_d^2E}{p_s^2(t)}\sum_{k\in\mathcal{K}}\frac{\rho_k^2(t)}{|h_k^{\rm dl}(t)|^2}}{1-4\eta_{t}^2E^2L^2}\right.\nonumber\\
	&+\left.\frac{2\eta_{t}^2E^3\left(\mu^2+4\delta\right)+8\eta_{t}^2E^3\mathbb{E}\left[\|\nabla F(\mathbf{w}(t))\|_2^2\right]}{1-4\eta_{t}^2E^2L^2}\right)\nonumber\\
	=&\frac{2\eta_{t}^3E^3L^2}{1-4\eta_{t}^2E^2L^2}\left(\mu^2+4\delta\right)+\eta_{t}\delta E\left(\sum_{k=1}^K\frac{1}{q_k}+1\right)\nonumber\\
	&+\frac{2d\sigma_d^2\eta_{t}EL^2}{(1-4\eta_{t}^2E^2L^2)p_s^2(t)}\sum_{k\in\mathcal{K}}\frac{\rho_k^2(t)}{|h_k^{\rm dl}(t)|^2}\nonumber\\
	&+\frac{20\eta_{t}^3E^3L^2-\eta_{t}E}{2-8\eta_{t}^2E^2L^2}\mathbb{E}\left[\|\nabla F(\mathbf{w}(t))\|_2^2\right],
\end{align}
where $(a)$ follows Lemma \ref{lem:bounded_SGD_iteration}.

Finally, combining \eqref{proof:eq2}, \eqref{eq:A2_bound_final}, and \eqref{eq:A1_bound_final}, we obtain
\begin{small}
	\begin{align}\label{proof:eq7}
		&\mathbb{E}\left[F(\mathbf{w}(t+1))\right]\nonumber\\
		\leq&\mathbb{E}\left[F(\mathbf{w}(t))\right]\nonumber\\
		&+\left(\frac{\eta_{t}E(20\eta_{t}^2E^2L^2+16\eta_{t}EL-1)}{2-8\eta_{t}^2E^2L^2}\right)\mathbb{E}\left[\|\nabla F(\mathbf{w}(t))\|_2^2\right]\nonumber\\
		&+\left(\frac{2\eta_{t}^2E^2L(1+\eta_tEL)}{1-4\eta_{t}^2E^2L^2}\right)\left(\mu^2+4\delta\right)+\eta_{t}\delta E\left(\sum_{k=1}^K\frac{1}{q_k}+1\right)\nonumber\\
		&+\left(\frac{d\sigma_d^2L(1+2\eta_tE+4\eta_t^2E^2L^2)}{(1-4\eta_{t}^2E^2L^2)p_s^2(t)}\right)\sum_{k\in\mathcal{K}}\frac{\rho_k^2(t)}{|h_k^{\rm dl}(t)|^2}\nonumber\\
		&+\frac{2dL}{\left(\sum_{j\in\mathcal{K}}a_j(t)p_j(t)\right)^2}.
	\end{align}
\end{small}
By subtracting $F^*$ at both sides of \eqref{proof:eq7}, we have
\begin{small}
	\begin{align}\label{proof:eq8}
		&\mathbb{E}\left[F(\mathbf{w}(t+1))-F^*\right]\nonumber\\
		\leq&\mathbb{E}\left[F(\mathbf{w}(t))-F^*\right]\nonumber\\
		&+\left(\frac{\eta_{t}E(20\eta_{t}^2E^2L^2+16\eta_{t}EL-1)}{2-8\eta_{t}^2E^2L^2}\right)\mathbb{E}\left[\|\nabla F(\mathbf{w}(t))\|_2^2\right]\nonumber\\
		&+\left(\frac{2\eta_{t}^2E^2L(1+\eta_tEL)}{1-4\eta_{t}^2E^2L^2}\right)\left(\mu^2+4\delta\right)+\eta_{t}\delta E\left(\sum_{k=1}^K\frac{1}{q_k}+1\right)\nonumber\\
		&+\left(\frac{d\sigma_d^2L(1+2\eta_tE+4\eta_t^2E^2L^2)}{(1-4\eta_{t}^2E^2L^2)p_s^2(t)}\right)\sum_{k\in\mathcal{K}}\frac{\rho_k^2(t)}{|h_k^{\rm dl}(t)|^2}\nonumber\\
		&+\frac{2d\sigma_u^2L}{\left(\sum_{j\in\mathcal{K}}a_j(t)p_j(t)\right)^2}\nonumber\\
		\overset{(a)}{\leq}&\left(1+\frac{\eta_{t}EL(20\eta_{t}^2E^2L^2+16\eta_{t}EL-1)}{1-4\eta_{t}^2E^2L^2}\right)\mathbb{E}\left[F(\mathbf{w}(t))-F^*\right]\nonumber\\
		&+\left(\frac{2\eta_{t}^2E^2L(1+\eta_tEL)}{1-4\eta_{t}^2E^2L^2}\right)\left(\mu^2+4\delta\right)+\eta_{t}\delta E\left(\sum_{k=1}^K\frac{1}{q_k}+1\right)\nonumber\\
		&+\left(\frac{d\sigma_d^2L(1+2\eta_tE+4\eta_t^2E^2L^2)}{(1-4\eta_{t}^2E^2L^2)p_s^2(t)}\right)\sum_{k\in\mathcal{K}}\frac{\rho_k^2(t)}{|h_k^{\rm dl}(t)|^2}\nonumber\\
		&+\frac{2d\sigma_u^2L}{\left(\sum_{j\in\mathcal{K}}a_j(t)p_j(t)\right)^2}\nonumber\\
		=&A(t)\mathbb{E}\left[F(\mathbf{w}(t))-F^*\right]+G(t),
	\end{align}
\end{small}
with
\begin{equation}
	A(t)=1+\frac{\eta_{t}EL(20\eta_{t}^2E^2L^2+16\eta_{t}EL-1)}{1-4\eta_{t}^2E^2L^2},
\end{equation}
and
\begin{small}
	\begin{align}
		G(t)=&\left(\frac{2\eta_{t}^2E^2L(1+\eta_tEL)}{1-4\eta_{t}^2E^2L^2}\right)\left(\mu^2+4\delta\right)\nonumber\\
		&+\eta_{t}\delta E\left(\sum_{k=1}^K\frac{1}{q_k}+1\right)\nonumber\\
		&+\left(\frac{d\sigma_d^2L(1+2\eta_tE+4\eta_t^2E^2L^2)}{(1-4\eta_{t}^2E^2L^2)p_s^2(t)}\right)\sum_{k\in\mathcal{K}}\frac{\rho_k^2(t)}{|h_k^{\rm dl}(t)|^2}\nonumber\\
		&+\frac{2d\sigma_u^2L}{\left(\sum_{j\in\mathcal{K}}a_j(t)p_j(t)\right)^2},
	\end{align}
\end{small}
where $(a)$ follows Lemma \ref{lem:bounded_global_gradient}. Assume the FL algorithm terminates after $T$ rounds, given an initial global model $\mathbf{w}_1$, we carry out recursions as
\begin{align}\label{eq:proof_lema1}
		&\mathbb{E}[F(\mathbf{w}(T+1))]-F^*\nonumber\\
		\leq&A(T)\mathbb{E}\left[F(\mathbf{w}(T))-F^*\right]+G(T)\nonumber\\
		\leq&A(T)\left(A(T-1)\mathbb{E}\left[F(\mathbf{w}(T-1))-F^*\right]+G(T-1)\right)\nonumber\\
		&+G(T)\nonumber\\
		\leq&\cdots\nonumber\\
		\leq&\prod_{t=1}^TA(t)\mathbb{E}[F(\mathbf{w}_1)-F^*]+\sum_{t=1}^{T-1}\left(\prod_{i=t+1}^{T}A(i)\right)G(t)\nonumber\\
		&+G(T).
\end{align}
Thus, this completes the proof.

\end{document}